\documentclass[aps,twocolumn,prc,showpacs,preprintnumbers,
               nofootinbib,float,superscriptaddress,longbibliography]{revtex4-1}
\usepackage{dcolumn}% Align table columns on decimal point
\usepackage{bm}% bold math
\usepackage{amssymb}
\usepackage{float}
\usepackage[caption = false]{subfig}

\usepackage{footnote}

\usepackage{hyperref}
\usepackage{xcolor}
\usepackage{graphicx}% Include figure files

\usepackage{epsfig}
\usepackage{color}
\newcommand{\km}{K^{\!^-}\!\!({ \bar{u}} { s})}
\newcommand{\ph}{\phi({ s} { \bar{s}})}
\newcommand{\al}{\bar{\Lambda}(\bar{u}\bar{d}\bar{s})}
\newcommand{\pbar}{\bar{p}({ \bar{u}\bar{u}}{ \bar{d}})}
\newcommand{\ks}{\overline{\Xi}^{+}(\bar{d}\bar{s}\bar{s})}
\newcommand{\om}{{\Omega}^{-}(sss)}
\newcommand{\op}{\overline{\Omega}^{+}(\bar{s}\bar{s}\bar{s})}

\usepackage{tikz}
\usetikzlibrary{shapes.geometric, arrows}

\tikzstyle{startstop} = [rectangle, rounded corners, minimum width=3cm, minimum height=1cm,text centered, draw=black, fill=red!30]
\tikzstyle{io} = [trapezium, trapezium left angle=70, trapezium right angle=110, minimum width=3cm, minimum height=1cm, text centered, draw=black, fill=blue!30]
\tikzstyle{process} = [rectangle, minimum width=3cm, minimum height=1cm, text centered, draw=black, fill=orange!30]
\tikzstyle{decision} = [rectangle, minimum width=3cm, minimum height=1cm, text centered, draw=black, fill=green!30]
\tikzstyle{comment} = [rectangle, rounded corners, minimum width=3cm, minimum height=1cm,text left, text width=4.2cm, draw=black, fill=yellow!30]
\tikzstyle{arrow} = [thick,->,>=stealth]

\begin{document}
\title{
Testing the impact of electromagnetic fields on the directed flow of constituent quarks in heavy-ion collisions 
}

\author{\it Ashik Ikbal Sheikh}
\email{asheikh2@kent.edu}
\address {Department of Physics, Kent State University, Kent, OH 44242, USA}
\author{\it Declan Keane}
\email{keane@kent.edu}
\address {Department of Physics, Kent State University, Kent, OH 44242, USA}

\author{\it Prithwish Tribedy}
\email{ptribedy@bnl.gov}
\address {Physics Department, Brookhaven National Laboratory, Upton, NY 11973, USA}

\date{\today}

\begin{abstract}
It has been proposed that strong electromagnetic fields produced in the early stages of heavy-ion collisions can lead to splitting of the rapidity-odd directed flow of positive and negative hadrons. For light hadrons, the interpretation of such measurements is complicated by the low magnitude of directed flow as well as by ambiguities arising from transported quarks. To overcome these complications, we propose measurements using only hadrons carrying
produced quarks ($\bar{u},\bar{d},s,\bar{s}$). We discuss how to identify the kinematics where such hadrons are produced via the  coalescence mechanism and therefore their flow is the sum of the flow of their constituent quarks. With this sum rule verified for certain combinations of hadrons, the expected systematic violation of this rule with increasing electric charge can be measured, which could be a consequence of the electromagnetic fields produced in the collisions. Our approach can be tested with the high statistics data from Phase II of the Beam Energy Scan (BES) program at the Relativistic Heavy Ion Collider (RHIC).
\end{abstract}

%\item{PACS numbers}
%\begin{keyword}


%\end{keyword}
%\end{frontmatter}

  \maketitle
%\pt{
\section{Introduction}

The first-order coefficient of azimuthal anisotropy for emitted particles, also known as directed flow, describes a collective sideward motion of particles in heavy-ion collisions~\cite{Voloshin:1994mz,Poskanzer:1998yz}. The rapidity-odd component of directed flow $v_1(y)$ (henceforth referred to as directed flow) has been argued to be sensitive to strong electromagnetic fields 
generated by the motion of the incoming protons in the colliding ions~\cite{Gursoy:2018yai,Gursoy:2014aka,Dubla:2020bdz}. 
As the spectator protons recede from the collision zone the produced magnetic field decays with time. This time-varying magnetic field induces an electric field due to the Faraday effect. The receding spectator protons also exert an electric force on the produced charged plasma in the collisions due to the Coulomb interaction. Another important effect comes into play as the medium produced in the collisions has an initial longitudinal expansion parallel to the beam direction and perpendicular to the direction of the magnetic field. 
The Lorentz force on the charged constituents results in an electric current perpendicular to both their velocity along expansion direction and magnetic field direction as a result of the Hall effect. If the combination of the Faraday and Coulomb effects 
%% wins over 
is stronger than 
the Hall effect, the directed flow of positively charged particles will become negative at positive rapidity ($v_1(h^+,y>0)<0$), and the opposite will happen for the negatively charged particles ($v_1(h^-,y>0)>0$). 
In other words, EM fields are expected to drive positively-charged and negatively-charged particles in opposite ways, %% eventually 
leading to a splitting of  $v_1(y)$~\cite{Gursoy:2018yai,Gursoy:2014aka}. 
This splitting is expected to be seen for various opposite-charge hadron pairs such as $\pi^\pm$, $K^\pm$, protons and antiprotons, etc., as calculated in Ref.~\cite{Gursoy:2018yai,Voronyuk:2014rna,Toneev:2016bri,Oliva:2019kin}. The strength of opposite-charge splitting of $v_1(y)$ depends on the magnitude of the electromagnetic fields, the lifetime of these fields, as well as on the electric charge and mass of the hadrons.

What are the factors that affect the observability of this splitting in experiment? First of all, it is most convenient to measure $v_1(y)$ with respect to the plane determined by the spectator deflection, as electromagnetic fields are largely generated by spectators. 
In a vacuum, the electromagnetic fields created by spectator protons increase in strength with beam energy, while their lifetime decreases with beam energy~\cite{Kharzeev:2007jp,Skokov:2009qp}. 
This scenario becomes complicated in the presence of a conducting medium. The charge-to-mass ratio of the colliding ions and the size of the system at a given centrality also leads to further complications ~\cite{Bzdak:2011yy,Deng:2012pc,Bloczynski:2012en,Tuchin:2013ie,Bloczynski:2013mca,McLerran:2013hla,Sun:2019hao}. 
The best choice is to perform a beam energy scan and measure $v_1$ at different centralities. As discussed later, the RHIC Beam Energy Scan (BES) program offers an ideal opportunity for this study. 

The optimum choice of particles to test the impact of electromagnetic fields using data from the RHIC BES program is the central theme of this paper. It was proposed by Das {\it et al.}~\cite{Das:2016cwd} that the measurement of $v_1$ splitting between particles with charm and anti-charm offers an advantage over particles with light flavor. This advantage arises from the fact that charm quarks are produced early in the collision and therefore experience the full early-stage electromagnetic fields as opposed to the weaker fields that die down with the evolution of the system. 
The proposal by Das {\it et al.}~\cite{Das:2016cwd} was to study $D^0\,(c\bar{u})$ and $\bar{D^0}\,(\bar{c}u)$ with the implicit assumption that while both these hadrons are neutral, the directed flow developed at the quark level is dominated by $c$ and $\bar{c}$ and drives the splitting of $v_1$.   
The $v_1$ splitting between $D^0$ and $\bar{D^0}$ mesons has been explored by the STAR~\cite{Adam:2019wnk} and ALICE~\cite{Acharya:2019ijj} collaborations. Due to the greater masses of heavy-flavor particles, the yield of these species is much lower than that of light-flavor particles and hence the measurements suffer from large statistical uncertainties. The problem of lower yields for heavy-flavors becomes acute if it is desired to study the energy dependence of such measurements, e.g., using the data from the second phase of RHIC BES program (BES-II), which was conducted during the years 2019-2021. Also, the measurements of $D^0$ and $\bar{D^0}$ at 200 GeV can not be repeated using the BES-II data already collected by the STAR collaboration due to the removal of the heavy flavor tracker (HFT)~\cite{starpsn0600} after the year 2016. 

Light hadron measurements offer several advantages since 
they are produced in abundance and have already been used to test splitting of $v_1$ driven by electromagnetic fields. The earliest such measurements from STAR reported $v_1$ splitting between positively and negatively charged hadrons in Cu+Au and Au+Au collisions at the top RHIC energy~\cite{Adamczyk:2016eux}. A significantly larger splitting is seen in asymmetric Cu+Au, and is attributed to the larger Coulomb force compared to symmetric Au+Au collisions. ALICE measurements for inclusive charged hadrons indicate a splitting of $v_1$ slope with about 2.6 $\sigma$ significance as a function of pseudorapidity, albeit weaker than that of the charmed hadrons ($D/\bar{D}$) measurements for which a significance of 2.7 $\sigma$ was observed \cite{Acharya:2019ijj}. 

Going below top RHIC energy, comprehensive measurements of $v_1$ for different light flavor particles and anti-particles have been reported in Refs.~\cite{Adamczyk:2014ipa} and \cite{Adamczyk:2017nxg} using BES-I data. However, interpretation of $v_1$ splitting using light-flavor particles in terms of electromagnetic fields runs into difficulties which can be avoided using the proposed method of the present paper.

\section{Method}
\label{method}

In the following discussion, we make two simplifying assumptions in a specific kinematic region: 
(1) coalescence of quarks is the dominant mechanism of hadron production, and (2) $v_1$ of a hadron is the sum of $v_1$ of its constituent quarks. 
The second assumption largely follows from the first~\cite{Adamczyk:2017nxg}.
These constituent quarks can either be produced in the collisions or transported from the incoming nuclei. Many of the emitted particle species are composed of constituent $u$ and $d$ quarks which might or might not be transported from the incoming nuclei \cite{Dunlop:2011cf}.
But the most important point for our discussion is as follows. The $v_1$ of transported quarks is quite different from that of the produced quarks ($\bar{u}$, $\bar{d}$, $s$ and $\bar{s}$)~\cite{Guo:2012qi}. This difference is due in part to 
the fact that a transported quark undergoes more interactions than a produced quark
\cite{Dunlop:2011cf,Adamczyk:2017nxg}. 
This in turn leads to a difference between $v_1$ of a hadron containing $u$ or $d$ quarks that could be either produced or transported, and $v_1$ of a hadron containing anti-quarks (that can only be produced), and greatly complicates \cite{Wang:2018pqx} the interpretation of $v_1$ splitting between positive and negative hadrons in terms of possible electromagnetic field effects. One specific example of the effect of transport on proton $v_1$ was studied using UrQMD simulation in Ref~\cite{Guo:2012qi}. In an experiment, one can measure directed flow of particles at midrapidity with respect to the plane determined by spectator deflection. This direction of projectile spectator deflection is conventionally taken as the positive $x$ direction~\cite{Voloshin:2016ppr}. In other words, the common convention is that projectile spectators that end up at positive beam rapidity have positive $v_1$ and target spectators that end up at negative beam rapidity have negative $v_1$~\cite{Voloshin:2016ppr}. UrQMD calculations~\cite{Guo:2012qi} at RHIC energies show that the transported protons have same direction of $v_1$ as the spectator nucleons and hence they have a positive $v_1$ slope ($dv_1/dy>0$) at mid-rapidity. On the other hand non-transported protons and anti-protons which can only be produced in the collisions are found to have $dv_1/dy<0$. This results in a positive splitting between protons and anti-protons, i.e., $\Delta dv_1/dy=dv_1/dy\,[p (uud)]- dv_1/dy\,[\bar{p}(\bar{u}\bar{u}\bar{d})]>0$. Based on such a study it is evident that transport will affect the splitting between any particle and anti-particle pairs having transported quark content, e.g., splitting between $\pi^+(u\bar{d})$ and $\pi^-(\bar{u}d)$; and also between $K^+ (u\bar{s})$ and $K^- (\bar{u}s)$. However, depending on the choice of particles and the quark content, the sign of the splitting maybe different and indistinguishable from EM-field-driven effects. 

 The complication arising from transported quarks can be bypassed by studying only particles such as $\km$, $\pbar$, $\al$, $\ph$, $\ks$, $\om$ and $\op$, which are entirely composed of produced constituent quarks, i.e., do not contain $u$ or $d$ quarks.
A natural question arises: these particles carry quarks with different flavor, electric charge ($q$) and mass ($m$) -- how to compare the $v_1$ of these seven different particle species and make a clean test of EM-field-driven effects? 
Given the known very strong dependence of directed flow on quark mass and flavor \cite{Adam:2019wnk, Acharya:2019ijj}, another factor to be considered is the different quark masses among the constituents of the selected hadron species. In other words, comparison between $v_1(\bar{d})$($q=1/3$) and $v_1(s)$($q=-1/3$) with an anticipation of EM-field-driven splitting due to a relative charge difference $\Delta q=2/3$ will be difficult to interpret due to the mass difference between $\bar{d}$ and s.

Unlike prior studies with a particle and an anti-particle, there is a difficulty in comparing the directed flow of the seven hadron species $\km$, $\pbar$, $\al$, $\ph$, $\ks$, $\om$ and $\op$. It is evident that there is no obvious way to compare two hadrons with similar constituent quark masses but different electric charge except in the case of $\Omega^{-}$ and $\overline{\Omega}^{+}$. Therefore our idea is to come up with useful combinations that will help us study directed flow splitting with increasing electric charge difference $\Delta q$.

 The first step in our approach is to select a kinematic region where our aforementioned assumption of the coalescence sum rule~\cite{Adamczyk:2017nxg} can be tested in a novel way, i.e., the directed flow of a suitably-chosen hadron species is consistent with the sum of the directed flow of its constituent quarks. 
In other words, we want to test if
\begin{eqnarray}
v_1({\rm hadron}) = \sum\limits_i v_1(q_i),
\label{eq_csr}
\end{eqnarray}
where the sum runs over the $v_1$ for the two constituent quarks $q_i$ in a meson and three in a baryon.

\begin{figure}
    \centering
    \includegraphics[width=0.5\textwidth]{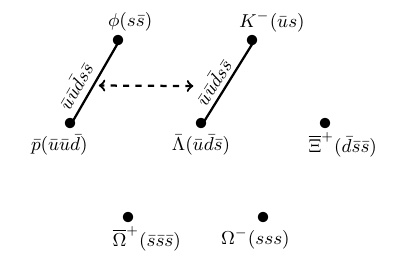}
    \caption{Diagram showing the seven abundantly produced hadrons composed of only produced (non-transported) quarks in heavy-ion collisions. We propose consideration of useful combinations such that the constituent quarks of two combinations to be compared at similar mass (assuming $m_u\sim m_d \ne m_s$) but at various values of relative electric charge $\Delta q$. The solid lines joining $\phi$ and $\bar{p}$ as well as between $K^-$ and $\bar{\Lambda}$ indicate that these two pairs have identical constituent quarks $\bar{u}\bar{u}\bar{d}s\bar{s}$ that correspond to the same mass $\Delta m=0$ and the same electric charge $\Delta q=0$. The sum of the directed flow of these two pairs must be similar if the coalescence sum rule holds ($v_1(\phi)+v_1(\bar{p})=v_1(K^-)+v_1(\bar{\Lambda})$) in the appropriate kinematic region. The geometric arrangement of the seven hadron species in this type of diagram has no special significance and is chosen simply to facilitate illustration of our chosen combinations.}
    \label{fig:cartoon_id1}
\end{figure}

Consistency with the coalescence sum rule can be investigated experimentally by testing the equality 

\begin{eqnarray}
v_1[\pbar] + v_1[\ph] = v_1[\km] + v_1[\al].
\label{eq_csr1}
\end{eqnarray}
Here, both left and right sides have the identical constituent quark content of $\bar{u}\bar{u}\bar{d}s\bar{s}$. However, the five quarks are distributed differently within the two pairs of hadrons. This is illustrated in Fig.~\ref{fig:cartoon_id1}. If $v_1$ at the level of constituent quarks is irrelevant, then Eq.\,(\ref{eq_csr1}) will not hold. However, if Eq.\,(\ref{eq_csr1}) holds in the appropriate kinematic region, this observation will support the sum rule. In such a case, one can also swap the anti-proton and the kaon in Eq.\,(\ref{eq_csr1}) to confirm that any apparent adherence to the sum rule is not due to an accidental balance of electric charge at the hadron level on both sides of the equality.

Alternatively, one can also test if 
%%DK if the following holds:\\
\begin{eqnarray}
v_1[\ks] + v_1[\km] = v_1[\al] + v_1[\ph].
\label{eq_csr2}
\end{eqnarray}
Both sides of Eq.\,(\ref{eq_csr2}) have the same quark content of $\bar{u}\bar{d}s\bar{s}\bar{s}$. If indeed the coalescence sum rule holds, we expect Eqs.\,(\ref{eq_csr1}) and (\ref{eq_csr2}) to hold at least in the appropriate kinematic region. 

 It is important to make sure that different terms in relations like Eqs.\,(\ref{eq_csr1}) and (\ref{eq_csr2}) are evaluated in a common region of rapidity [$y_{\rm min}\le y \le y_{\rm max}$] and transverse momentum per constituent quark [$(p_T/n_q)_{\rm min} \le p_T/n_q \le (p_T/n_q)_{\rm max}$, where $n_q$ is the number of constituent quarks]. The requirement for a common $p_T/n_q$ follows from the fact that in the coalescence mechanism, the transverse momentum, $p_T^{B}$ and $p_T^{M}$, of a baryon $B(q_1\,q_2\,q_3)$ and a meson $M(q\,\bar{q})$ respectively, is the sum of the transverse momentum of their constituent quarks. In other words,
\begin{eqnarray}
    v_1^{B}(p_T^{B}) &=& v_1^{q_1}(p_T^{B}/3) + v_1^{q_2}(p_T^{B}/3) + v_1^{q_3}(p_T^{B}/3) \\
    v_1^{M}(p_T^{M}) &=& v_1^{q}(p_T^{M}/2) +  v_1^{\bar{q}}(p_T^{M}/2). 
\end{eqnarray}
Therefore, in an experiment if one measures $v_1$ of anti-proton, $\phi$ meson, $K^-$ and $\bar{\Lambda}$ in $y_{\rm min}\le y \le y_{\rm max}$, as a function of their transverse momenta $p_T^{\bar{p}},~p_T^\phi,~p_T^{K^-}$\,and $p_T^{\bar{\Lambda}}$, respectively, 
Eq.\,(\ref{eq_csr1}) should be evaluated in the kinematic region where $(p_T/n_q)_{\rm min}< (p_T^{\bar{p}}/3), (p_T^\phi/2),  (p_T^{K^-}/2), (p_T^{\bar{\Lambda}}/3)<(p_T/n_q)_{\rm max}$.

Similar consistency with the coalescence sum rule is not expected to be satisfied when the particle species involved can include transported constituent quarks. This has been discussed previously \cite{Dunlop:2011cf} and has also been investigated using STAR data \cite{Adamczyk:2017nxg, Wang:2018pqx} and indeed a violation was observed. 
 In analogy with Eq.\,(\ref{eq_csr1}) and Eq.\,(\ref{eq_csr2}), a straightforward way to experimentally test the role of transported quarks is to observe violations of the two following  equalities

\begin{table*}[th]
\renewcommand{\arraystretch}{1.5}
\begin{tabular}{|l l l l l l l|}
\hline
Index & Quark mass && Charge & & Strangeness & ~~~~~~~$\Delta v_1$ combination   \\ 
\hline
1 & $\Delta m=0$ & & $\Delta q=0$ & & $\Delta S=0$ & ${[\pbar + \ph]}-[\km + \al ] $ \\ 
2 & $\Delta m=0$ & & $\Delta q=0$ & & $\Delta S=0$ & ${ [\ks + \km]}-[\al + \ph]$ \\
\hline
3 & $\Delta m\approx 0$ & & $\Delta q=\frac{1}{3}$ & & $\Delta S=0$ & ${ \frac{1}{3}[\om + \pbar]}-[\km]$ \\
\hline
4 & $\Delta m\approx 0$ & & $\Delta q=\frac{2}{3}$ & & $\Delta S=1$ & ${ [\al]}-[\frac{1}{2} \ph +\frac{2}{3} \pbar]$  \\
\hline
5 & $\Delta m\approx 0$ & & $\Delta q=1$ & & $\Delta S=2$ &
${[\al]}-[\frac{1}{3}\om + \frac{2}{3}\pbar]$\\
\hline
6 & $\Delta m\approx 0$ & & $\Delta q=\frac{4}{3}$ & & $\Delta S=2$ &
${[\al]}-[\km + \frac{1}{3}\pbar]$ \\
7 & $\Delta m\approx 0$ & & $\Delta q=\frac{4}{3}$ & & $\Delta S=2$ &
${[\ks]}-[\ph + \frac{1}{3}\pbar]$\\
\hline
8 & $\Delta m\approx 0$ & & $\Delta q=\frac{5}{3}$ & & $\Delta S=2$ &
${[\ks]}-[\km + \frac{1}{3}\op]$ \\
\hline
9 & $\Delta m= 0$ & & $\Delta q=2$ & & $\Delta S=6$ &
${[\op]}-[\om]$ \\
\hline
10 & $\Delta m\approx 0$ & & $\Delta q=\frac{7}{3}$ & & $\Delta S=4$ &
${[\ks]}-[\km + \frac{1}{3}\om]$ \\
\hline
\end{tabular}
\caption{Differences between combinations formed from hadron species composed of produced quarks only. In all cases, the constituent quark mass difference ($\Delta m$) is zero or near-zero, while the charge difference ($\Delta q$) and strangeness difference ($\Delta S$) is varied as tabulated. %Seven out of these ten equations are linearly independent (indices 1, 4, 5, 6, 7, 9 and 10). 
The ten expressions are not linearly independent. A set of linearly independent expressions can be found using linear algebra. See Appendix~\ref{a1:linal} for details. For visualization purpose, indices 1, 6 and 10 are illustrated in diagrammatic form in Figs.\,\ref{fig:cartoon_id1}, \ref{fig:cartoon_id2} and \ref{fig:cartoon_id3}, respectively.}
\label{tab:delq_dels}
\end{table*}
\begin{eqnarray}
\nonumber
v_1[\pi^{+}(u\bar{d})] + v_1[\pi^{-}(\bar{u}d)] + v_1[K^{+}(u\bar{s})] \\ = v_1[p(uud)] + v_1[\bar{\Lambda}(\bar{u}\bar{d}\bar{s})]
\label{eq_csr4}
\end{eqnarray}

\begin{eqnarray}
\nonumber
v_1[\pi^{+}(u\bar{d})] + v_1[\pi^{-}(\bar{u}d)] + v_1[K^{-}(\bar{u}s)] \\ = v_1[\bar{p}(\bar{u}\bar{u}\bar{d})] + v_1[\Lambda(uds)]
\label{eq_csr5}
\end{eqnarray}
Here, both sides of Eq.\,(\ref{eq_csr4}) and both sides of Eq.\,(\ref{eq_csr5}) have the identical quark content $uu\bar{u}d\bar{d}\bar{s}$ and $u\bar{u}\bar{u}d\bar{d}s$, respectively. In spite of this identical quark content, the presence of 
unknown fractions of transported quarks $u$ and $d$ will result in observed discrepancies. Even if a violation of Eq.\,(\ref{eq_csr4}) or Eq.\,(\ref{eq_csr5}) is observed, it is important to determine whether such a violation is due to a difference between the mass of the hadrons on both sides of Eq.\,(\ref{eq_csr4}) and Eq.\,(\ref{eq_csr5}). This question can be clarified if one observes the expected discrepancies in the equality using neutral $\Lambda$ hyperons: 
\begin{eqnarray}
v_1[\Lambda(uds)] = v_1[\al].
\label{eq_csr6}
\end{eqnarray}
A violation of Eq.\,(\ref{eq_csr6}) can not be due to a difference in the mass of hadrons or quarks, or due to a difference in the electric charge -- hence it can only be attributed to the role of transported quarks. 
Experimental investigation of discrepancies such as in Eqs.~(\ref{eq_csr4}), (\ref{eq_csr5}) and (\ref{eq_csr6}) are as important as observing the validity of Eqs.~(\ref{eq_csr1}) and (\ref{eq_csr2}). They can provide more insight into the transport of quarks from the initial nuclei, but this is not the focus of the present work. 

Having identified the appropriate region in $y$ and $p_T/n_q$ where Eqs. (\ref{eq_csr1}) and (\ref{eq_csr2}) are observed to be valid, we move on to test for possible and systematic violation of the sum rule in cases of non-identical constituent quark combinations, which is the main focus of this paper. 
For convenience of discussion, we express such combinations in terms of a difference $\Delta v_1$ and call this the ``splitting of $v_1$". For example, we re-write Eq.(\ref{eq_csr1}) as
\begin{eqnarray}
\nonumber
\Delta v_{1} (\Delta q=0,\, \Delta S=0) = ~~~~~~~~~~~~~~~~~~~~~\\ 
\{v_1[\pbar] + v_1[\ph]\} - \{v_1[\km] + v_1[\al]\},~
\label{delv1-id}
\end{eqnarray}
for which we expect $\Delta v_1(\Delta q=0,\, \Delta S=0) \approx0$. 
Here, $\Delta q$ and $\Delta S$ refer to the difference in electric charge and strangeness, respectively, for the above combinations. Similarly, one can study 
\begin{eqnarray}
\nonumber
&\Delta v_{1} (\Delta q=\frac{4}{3},\, \Delta S=2) = \\ 
& v_1[\al] - \{v_1[\km] + \frac{1}{3} v_1[\pbar]\},
\label{delv1-nonid}
\end{eqnarray} 
and 
\begin{eqnarray}
\nonumber
&\Delta v_{1} (\Delta q=\frac{7}{3},\, \Delta S=4) = \\ 
& v_1[\ks] - \{v_1[\km] + \frac{1}{3} v_1[\om]\}.
\label{delv1-nonid2}
\end{eqnarray}

\begin{figure}
    \centering
    \includegraphics[width=0.5\textwidth]{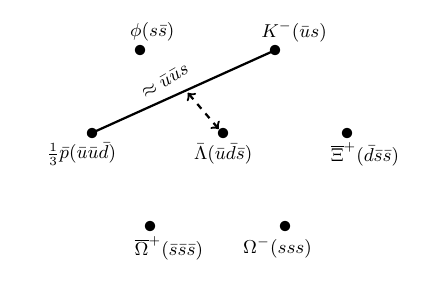}
\caption{A diagram similar to Fig.~\ref{fig:cartoon_id1}. The idea is to combine the $v_1$ of $K^-$ with one third of the $v_1$ of $\bar{p}$. Assuming quark masses $m_u\sim m_d \ne m_s (=m_{\bar{s}})$, the combination is approximately equivalent to the directed flow of the sum of $\bar{u}\bar{u}s$. This quantity can be compared to $v_1$ of $\bar{\Lambda}$ which has constituent quarks $\bar{u}\bar{d}\bar{s}$. See Index 6 in Table~\ref{tab:delq_dels}. This enables us to study possible splitting of directed flow at nearly same mass $\Delta m=2(m_{\bar{d}}-m_{\bar{u}})/3\approx 0$ but at relative electric charge of $\Delta q=4/3$, i.e. $\Delta v_1=v_1(\bar{\Lambda})-v_1(K^-)-v_1(\bar{p})/3$ which is expected to be nonzero due to EM-field-driven effects. Further expressions similar to this can be found in Table~\ref{tab:delq_dels}.} 
\label{fig:cartoon_id2}
\end{figure}

\begin{figure}
    \centering
    \includegraphics[width=0.5\textwidth]{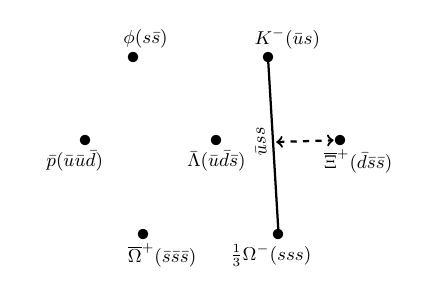}
\caption{A diagram similar to Fig.~\ref{fig:cartoon_id2}. Here the $v_1$ of $K^-$ is combined with one third of the $v_1$ of $\Omega^{-}$. The combination is equivalent to the directed flow of the sum of $\bar{u}ss$. This quantity can be compared to $v_1$ of $\overline{\Xi}^{+}$ which has constituent quarks $\bar{d}\bar{s}\bar{s}$. This also helps us to study possible splitting of directed flow at approximately same mass $\Delta m=m_{\bar{d}}-m_{\bar{u}}\approx 0$ but at relative electric charge of $\Delta q=7/3$, i.e. $\Delta v_1=v_1(\overline{\Xi}^{+})-v_1(K^-)-v_1(\overline{\Omega}^{+})/3$ which is expected to be nonzero due to EM-field-driven effects. See Index 10 in Table~\ref{tab:delq_dels}. More expressions similar to this are given in Table~\ref{tab:delq_dels}.} 
\label{fig:cartoon_id3}
\end{figure}
The splitting for the expressions above is illustrated diagrammatically in Figs.~\ref{fig:cartoon_id2} and~\ref{fig:cartoon_id3}. Note that these expressions are
designed to make sure that the sum of the bare masses of the quarks are similar on both sides of the minus sign, where we assume $m_u\sim m_d \ne m_s (=m_{\bar{s}})$. The net differences in the sum of bare masses in Eq.\,(\ref{delv1-nonid}) and Eq.\,(\ref{delv1-nonid2}) are $\Delta m=2(m_{\bar{d}}-m_{\bar{u}})/3$ and $\Delta m=m_{\bar{d}}-m_{\bar{u}}$ respectively, which are negligible. Equations (\ref{delv1-nonid}) and (\ref{delv1-nonid2}) have an imbalance in electric charge of $\Delta q=4/3$ and $\Delta q=7/3$, respectively. 
If electromagnetic fields play a role leading to electric charge dependence of $v_1$, one expects $\Delta v_1(\Delta q=4/3,\, \Delta S=2) \ne 0$ and $\Delta v_1(\Delta q=7/3,\, \Delta S=4) \ne 0$. Many relations like Eqs.~(\ref{delv1-id}), (\ref{delv1-nonid}) and (\ref{delv1-nonid2}) can be constructed.

\begin{figure*}
   \centering
    \includegraphics[width=0.86\textwidth]{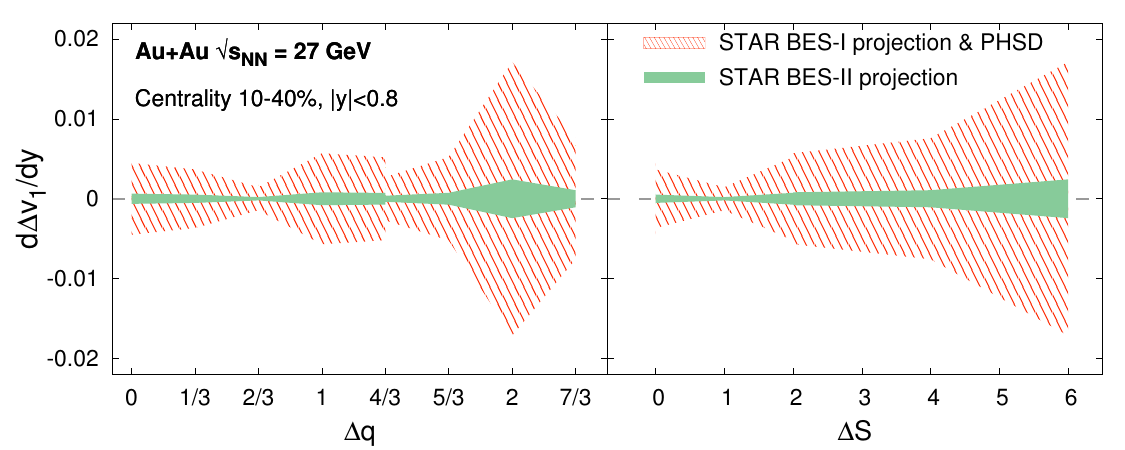}
    \caption{
    (Left): Expected uncertainties in $\Delta v_1$ slope ($d\Delta v_1/dy$) for STAR BES-I and BES-II data sets near mid-rapidity ($|y|<0.8$), as a function of electric charge difference $\Delta q$ (see Table~\ref{tab:delq_dels}). Other selections assumed here include 10-40\% centrality in Au+Au collisions at $\sqrt{s_{NN}} = 27$ GeV. (Right): The same as the left panel, but as a function of strangeness difference $\Delta S$ (see Table~\ref{tab:delq_dels}). Events generated using the PHSD microscopic transport approach are used here~\cite{Bratkovskaya2011PartonHadronStringDA, Konchakovski:2014gda,Palmese:2016abq}.
   }
  \label{fig:slope_delq_dels}
\end{figure*}

The anticipated useful combinations using the seven hadron species $\km$, $\pbar$, $\al$, $\ph$, $\ks$, $\om$ and $\op$, each produced with at least adequate abundance at RHIC/BES energies, are listed in Table~\ref{tab:delq_dels}. Although Table~\ref{tab:delq_dels} shows ten expressions, only five of them are linearly independent. 

Among listed combinations in Table~\ref{tab:delq_dels}, indices 1 and 2 have identical quark content on both sides of the minus sign, i.e., $\Delta q =\Delta S =0$, and therefore are most likely to adhere closely to the coalescence sum rule and yield a measurement $\Delta v_1 =0$.
The remaining combinations, indices 3-10, provide a way to study $\Delta v_1$ with an increasing value of $\Delta q$. Since the  splitting of directed flow due to electromagnetic fields is stronger with a larger electric charge difference, one expects stronger deviation of $\Delta v_1$ from zero with increasing $\Delta q$. 
The advantage of our approach over previous measurements of directed flow splitting, namely using a pair of positive and negative hadrons, should be noted here. Since most hadrons are pions, previous measurements correspond to the study of $\Delta v_1$ at the fixed value $\Delta q=2$. Our approach provides a large lever-arm to extend the investigation over a range of $\Delta q$ from 0 to $7/3$ using our seven selected hadron species. An additional advantage is that one avoids the effects of transported quarks. 

Table~\ref{tab:delq_dels} indicates that there are degenerate combinations, i.e., combinations 1 and 2 as well as combinations 6 and 7. These degenerate combinations are useful to cross-check the results. Since we start with measurements of seven hadron species, the ten combinations shown in Table~\ref{tab:delq_dels} are not linearly independent. By employing linear algebra (discussed in Apendix~\ref{a1:linal}), possible sets of independent combinations can be identified. One possible choice of five linearly independent expressions is 4, 5, 6, 9 and 10. This is not a unique choice. Other two possible choices are 1, 4, 5, 9 and 10; also 1, 5, 6, 9 and 10. Also, by adding more hadrons, one can construct more expressions than shown in Table~\ref{tab:delq_dels}. Measurements of hadrons more massive than $\Omega$ will be challenging at RHIC/BES energies, but can be explored at higher energies. The quantity $\Delta v_1$ can be studied as a function of rapidity. In this way, one can estimate the slope $d\Delta v_1/dy$ over the range of rapidity allowed by the experiment. The dependence of $d\Delta v_1/dy$ on $\Delta q$ will provide an important insight into the role of EM-field-driven effects. One expects increasingly positive or negative values of $d\Delta v_1/dy$ with increasing $\Delta q$ as evidence of EM-field-driven effects depending on the dominance among the Faraday and Hall effects.

It is clear from Table~\ref{tab:delq_dels} that the discussed increase in $\Delta q$ is also associated with a change in $\Delta S$. This is an unavoidable consequence of the quantum numbers carried by the constituent quarks. A mechanism that leads to splitting of directed flow between strange and anti-strange quarks is not known. 
If such a mechanism is identified, it may complicate any conclusion from the $\Delta q$ dependence of $\Delta v_1$. A possible way to disentangle the effects of $\Delta q$-driven separation and $\Delta S$-driven separation of $\Delta v_1$ is as follows. 
As shown in Table~\ref{tab:delq_dels}, there are four combinations (indices 5-8) that correspond to the same $\Delta S$ (=2) but different $\Delta q$. These combinations have a special importance in the sense that they can be employed to measure the $\Delta v_1$ slope as a function of $\Delta q$ for the same $\Delta S$ (=2). Such measurements would be helpful for further understanding the effect of the electromagnetic field on the splitting. However, the challenge for this measurement is that the lever-arm for variation of $\Delta q$ will be small and restricted to the range of $\Delta q=1$ to $\Delta q=5/3$. Such a measurement will require high precision. In any case, if a systematic deviation of $\Delta v_1$ with $\Delta S$ is observed, such an observation will also be interesting and will deserve further investigation. 

\section{Projection for RHIC-BES program}
\label{bes-proj}
%\pt{Ashik, please add citations wherever needed, take a look at the isobar paper, it has all the references}

The proposed approach can be tested experimentally using the upcoming large BES-II data samples from the STAR experiment in a model-independent way. Model calculations including effects of electromagnetic fields using a simulation framework such as iEBE-VISHNU~\cite{Gursoy:2018yai} or Parton-Hadron String Dynamics (PHSD)~\cite{Voronyuk:2011jd, Bratkovskaya2011PartonHadronStringDA, Konchakovski:2014gda, Palmese:2016abq} would be ideal to test our approach. However, such calculations are beyond the scope of the present paper. Here the main goal is to explain the proposed analysis approach, while a secondary objective is to estimate the likely statistical uncertainties for upcoming BES-II measurements from the STAR experiment for Au+Au collisions at a typical energy, namely $\sqrt{s_{NN}}=27$ GeV. For this, we use  published BES-I measurements~\cite{Adamczyk:2014ipa,Adamczyk:2017nxg} as well as a set of events generated using the PHSD microscopic transport approach~\cite{PHSDprivate}.
We choose the PHSD model as it provides reasonable descriptions of the rapidity dependence of $v_1$ for $K^\pm$, $p/\bar{p}$ and $\Lambda (\bar{\Lambda})$ in 10-40\% Au+Au collisions at $\sqrt{s_{NN}}=27$ GeV~\cite{Soloveva:2020ozg}. We focus on this centrality bin for our estimates.  
The full PHSD codes are available by registration, but PHSD simulations are computationally intensive. With the help of PHSD authors, we generated about 2 million PHSD events for the purpose of making our projection plots. The PHSD simulation was performed for Au+Au collisions at $\sqrt{s_{NN}}=27$ GeV with minimum (maximum) impact parameter $b_{\rm min} (b_{\rm max})$ = 2.0 (8.7) fm, corresponding to 10-40\% centrality. The model simulation incorporated a partonic QGP phase, and the final time step of the calculation was $\tau_{F}$ = 150 fm/$c$. 

So far, the available $v_1$ measurements from BES-I include four particle species composed of produced constituent quarks only~\cite{Adamczyk:2014ipa, Adamczyk:2017nxg}: $K^-$, $\bar{p}$, $\bar{\Lambda}$ and $\phi$. Measurements of $v_1$ for $\Xi$ and $\Omega$ baryons have not been reported at RHIC, and for these species, PHSD model calculations are essential.  
Using the available measurements from BES-I for 10-40\% central Au+Au collisions at $\sqrt{s_{NN}}=27$ GeV over the rapidity window $|y|<0.8$ ~\cite{Adamczyk:2014ipa,Adamczyk:2017nxg}, we can estimate $d\Delta v_1/dy$ for $\Delta q = 0$, $1/3$ and $4/3$, and for $\Delta S = 0$, $1$ and $2$.  These correspond to indices 1, 4 and 6 in Table~\ref{tab:delq_dels} and do not include $\Xi$ or $\Omega$. The magnitudes of $d\Delta v_1/dy$ from BES-I data, which were analyzed for only 70 M events, do not show any significant deviation from zero for $\Delta q = 0$, $1/3$ or $4/3$. We therefore only focus on the uncertainties related to these measurements. 

We use 2 million events from the PHSD model to estimate the same quantities in 10-40\% central Au+Au $\sqrt{s_{NN}}=27$ GeV collisions and obtain the ratio of uncertainties between BES-I data and PHSD calculations for different quantities. These ratios range from 0.334 to 2.03 due to various factors such as the difference in the event numbers between data and the PHSD simulation, the experimental correction factors to incorporate resolution of the first-order harmonic event plane, and the uncertainty due to acceptance and reconstruction of different particle species. We use the maximum value of this ratio to scale the uncertainties for the quantity $d\Delta v_1/dy$ for the remaining relations involving $\Xi$ and $\Omega$ shown in Table~\ref{tab:delq_dels} obtained from PHSD model. The results obtained by this method are presented as red hatched bands in Fig.\,\ref{fig:slope_delq_dels}, where we plot the uncertainty on $d\Delta v_1/dy$ as a function of $\Delta q$. The mean values of the quantities are set to zero. It is clear from Fig.\,\ref{fig:slope_delq_dels} that uncertainties are large when combinations involve the heavier species, but will be significantly  improved when BES-II data become available for analysis. 

The STAR collaboration has collected large samples of Au+Au collisions at 11 beam energies in collider mode, and at 11 partly overlapping beam energies in fixed-target mode during the period 2019 to 2021~\cite{StarBur:2021}. In the current paper, we illustrate the projected improvement in statistical errors at the representative energy point of $\sqrt{s_{NN}}=27$ GeV. 
Based on the number of events used in BES-I ($N_{\rm evt}^{\rm BES-I}$) measurements and the number of collected events in BES-II ($N_{\rm evt}^{\rm BES-II}$), we obtain another scale factor,
\begin{eqnarray}
k = C_{\rm EP}^{\rm Res} \sqrt{\Big(\frac{N_{\rm evt}^{\rm BES-II}}{N_{\rm evt}^{\rm BES-I}} \Big)}.
\end{eqnarray}
Here we use the factor ($C_{\rm EP}^{\rm Res}$) to account for the high event-plane resolution provided by the STAR Event Plane Detector (EPD)~\cite{Adams:2019fpo} since 2018, relative to the previously used Beam-Beam Counter (BBC)~\cite{Whitten:2008zz}. 
For Au+Au collisions at $\sqrt{s_{NN}}=27$ GeV and 10-40\% centrality, we obtain $k=3.48$ with $N_{\rm evt}^{\rm BES-I} = 70$M~\cite{Adamczyk:2017nxg}, $N_{\rm evt}^{\rm BES-II} =400$M~\cite{StarBur:2021} and $C_{\rm EP}^{\rm Res} = 1.45$~\cite{Adams:2019fpo} (for mid-central collisions). Then the uncertainty projections for $d\Delta v_1/dy$ in BES-II, estimated by scaling the corresponding numbers for BES-I with the obtained $k$ factor, are shown by solid green bands in Fig.\,\ref{fig:slope_delq_dels}. 
%}

\begin{figure}[htb]
   \centering
    \includegraphics[width=0.5\textwidth]{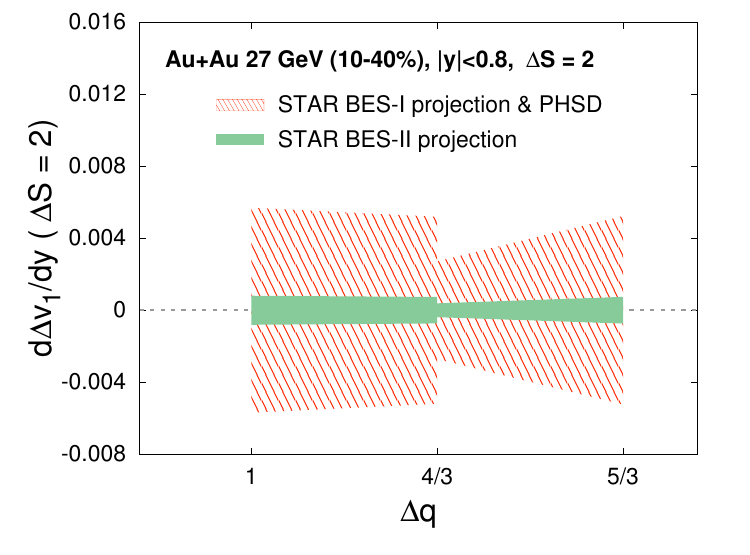}
    \caption{
    For the condition $\Delta S=2$, the expected uncertainties in $d\Delta v_1/dy$ for STAR BES-I and BES-II data sets near mid-rapidity ($|y|<0.8$), as a function of electric charge difference $\Delta q$. Other selections assumed here include 10-40\% centrality in Au+Au collisions at $\sqrt{s_{NN}} = 27$ GeV.
    }
  \label{fig:slope_delq_dels-2}
\end{figure}

Figure~\ref{fig:slope_delq_dels-2} presents expected uncertainties in $d\Delta v_1/dy$ at a fixed $\Delta S$ of 2 for the case of STAR BES-I and BES-II. These calculations apply to $d\Delta v_1/dy$ near mid-rapidity ($|y| < 0.8$) as a function of electric charge difference ($\Delta q$) for 10-40\% centrality in Au+Au collisions at $\sqrt {s_{NN}} = 27$ GeV. 

Although our projected statistical errors motivate the proposed measurements based on available BES-II data, caveats remain for our anticipated uncertainties related to $\Xi$ and $\Omega$, for which $v_1$ measurements have not reported, and where we have used PHSD to make conservative estimates. We have ignored the fact that because these hyperon species are reconstructed via their weak decays, their measured flow is subject to uncertainties related to reconstruction and background contamination~\cite{STAR:2019bjj}. However, the STAR collaboration has published global polarization results for these hyperon species ~\cite{STAR:2020xbm} and has demonstrated that such uncertainties can be minimized. In the case of $\Xi$ and $\Omega$ using the Kalman Filter reconstruction method~\cite{Zyzak:Phd,Gorbunov:Phd,Kisel:2018nvd}, a purity of $90\%$ is achieved, which is about $30\%$ better than the traditional method of $\Xi$ and $\Omega$ reconstruction previously used for BES-I data~\cite{STAR:2019bjj}. These improvements will be crucial to incorporate in the upcoming measurements from BES-II. Despite the caveats, our error projections shown in Fig.\,\ref{fig:slope_delq_dels} and \ref{fig:slope_delq_dels-2} indicate that BES-II data provide an ideal opportunity to test the proposed approach. In addition, the inner Time Projection Chamber (iTPC) upgrade of the STAR detector will enhance the pseudorapidity acceptance from $|\eta|<1$ to $|\eta|<1.5$ 
%% and therefore allow studies of $v_1$ over a wider range of $y$
~\cite{STARnote:644}. 

While discussions here have focused on RHIC/BES energies, our approach can be used at the top RHIC energy and at the LHC energies.  %\textcolor{red}{Nevertheless, it is noteworthy to mention that the approach is based on the naive coalescence mechanism and a departure from the proposed pattern might be observed at the lower beam energies. It has been seen before at STAR, the underlying assumptions of coalescence picture breaks down between beam energy 11.5 and 7.7 GeV~\cite{Adamczyk:2017nxg}. We also point out that the contamination due to transported quarks has a collision energy dependence and it is suppressed at higher energies, like the LHC. Regardless of whether the contamination is greatly suppressed or the coalescence sum rule holds at the higher energies, it will be of interest to study the difference between the directed flow of $\Omega^-$ and $\bar{\Omega}^+$ at the LHC where the abundance is much larger than at the RHIC.}
The present approach involves a quark coalescence mechanism, and in a scenario where the beam energy drops below the onset of production of a quark-gluon plasma phase, the proposed pattern might change or might no longer be observed. The prerequisite for such pattern is expected to be dependent on the centrality of collisions and the size of the colliding nuclei. It has been reported by the STAR collaboration, based on the relatively poor statistics of BES-I, that the index 6 difference, shown on Table ~\ref{tab:delq_dels}, is zero within errors between 200 and 11.5 GeV but its magnitude increases sharply at 7.7 GeV~\cite{Adamczyk:2017nxg}. We also point out that the contamination due to transported quarks has a collision energy dependence and it is suppressed at higher energies, like at the LHC. Regardless, it will be of interest to study the difference between the directed flow of $\Omega^-$ and $\bar{\Omega}^+$ at the LHC where the abundance is much larger than at RHIC.

It will also be interesting to test our approach using the large isobar collision data set from RHIC that has been recently studied in the context of the Chiral Magnetic Effect (CME)~\cite{STAR:2021mii}.  Although no signature of CME is observed, it is still of interest to exploit the difference in the $B$-field between Ru+Ru and Zr+Zr to study splitting of $v_1$. Our approach can also be applied to study the effect of EM fields~\cite{Gursoy:2018yai, Dunlop:2011cf, Goudarzi:2020eoh} in higher-order harmonic flow coefficients.

\section{Summary}
\label{summary}
 The effect of EM fields is expected to lead to a splitting between the rapidity dependence of the directed flow of hadrons carrying different electric charges. Prior studies to test such predictions using a pair of light hadrons like $\pi^\pm$ or $K^\pm$ have major drawbacks such as contamination due to transported quarks. Also, such approaches study directed flow splitting $\Delta v_1$ at a fixed value of electric charge difference $\Delta q=2$ between the hadron pairs, and therefore have limited lever-arm to enhance the possible signature of EM-field-driven effects. 
 To overcome these problems, we propose a new analysis method based on measurement of directed flow of several particle species composed of produced constituent quarks only, namely $K^-$, $\bar{p}$, $\bar{\Lambda}$, $\phi$, $\overline{\Xi}^-$, ${\Omega}^-$ and $\overline{\Omega}^+$ to test the effect of EM fields in heavy-ion collisions.
In this approach, particle species are combined such that comparison of the directed flow can be performed at the constituent quark level at various values of electric charge but at nearly equal mass. 
We first demonstrate that combinations of the aforementioned particle species are well suited to test that the flow of a hadron equals the sum of the flow of its constituent quarks (the coalescence sum rule) when identical quark combinations are compared. 
Once a kinematic region is identified where the sum rule holds, the next step is to study the deviations in directed flow, i.e., the splitting $\Delta v_1$, with increasing values of $\Delta q$. 
We discuss linearly independent relations obtained by combining the seven above-mentioned hadron species in such a way that a wide lever-arm is obtained to vary $\Delta q$ from 0 to $7/3$ and measure $\Delta v_1$. 

Large data samples from the second phase of the Beam Energy Scan program at the STAR experiment at Brookhaven National Laboratory can be used to measure the proposed observables, and ought to be precise enough to gain new information on possible effects of the strong early-stage electromagnetic fields in heavy-ion collisions. The inclusion of multi-strange hadrons in this type of analysis opens up further opportunities.

\section*{Acknowledgment}
We are thankful to Elena Bratkovskaya, Lucia Oliva, Vadim Voronyuk and Olga Soloveva for assistance with the PHSD model. We thank Chun Shen for useful discussions. A.I.S. and D.K. acknowledge support from the Office of Nuclear Physics within the US DOE Office of Science, under Grant DE-FG02-89ER40531. P.T. is supported under DOE Contract DE-SC0012704.

\appendix
\section{ Evaluation of linearly independent combinations
from Table~\ref{tab:delq_dels}}
\label{a1:linal}

\begin{table*}[tbh]
\renewcommand{\arraystretch}{1.5}
\begin{tabular}{|l l l l l|}
\hline
Index & & ~~~~~ Expression & & ~~~~~~~~~~~~~Vector  \\ 
\hline
1 && ${\pbar + \ph}-\km - \al $ & & $v_1=\{-1,1,-1,1,0,0,0\}$ \\ 
2 && ${\ks + \km} - \al - \ph$ & & $v_2=\{1,0,-1,-1,1,0,0\}$ \\ 
\hline
3 && ${ \frac{1}{3}\om + \frac{1}{3}\pbar}-\km$ && $v_3=\{-1,1/3,0,0,0,1/3,0\}$ \\
\hline
4 && ${\al}-\frac{1}{2} \ph -\frac{2}{3} \pbar$  && $v_4=\{0,-2/3,1,-1/2,0,0,0\}$ \\
\hline
5 && ${\al}-\frac{1}{3}\om - \frac{2}{3}\pbar$ && $v_5=\{0,-2/3,1,0,0,-1/3,0\}$ \\
\hline
6 && ${\al}-\km - \frac{1}{3}\pbar$ && $v_6=\{-1,-1/3,1,0,0,0,0\}$ \\
7 && ${\ks}-\ph - \frac{1}{3}\pbar$ && $v_7=\{0,-1/3,0,-1,1,0,0\}$ \\
\hline
8 && ${\ks}-\km - \frac{1}{3}\op$ && $v_8=\{-1,0,0,0,1,0,-1/3\}$ \\
\hline
9 && ${\op}-\om$ && $v_9=\{0,0,0,0,0,-1,1\}$ \\
\hline
10 && ${\ks}-\km - \frac{1}{3}\om$ && $v_{10}=\{-1,0,0,0,1,-1/3,0\}$ \\
\hline
\end{tabular}
\caption{Table for the vectors constructed from each combination in Table ~\ref{tab:delq_dels}. The vectors are formulated in the {\bf R}$^7$ vector space where the basis is formed by the seven particle species discussed here, namely, $K^{-}, \bar{p},\, \bar{\Lambda},\, \phi,\, \overline{\Xi}^{+}, \Omega^{-}, \overline{\Omega}^{+}$.}
\label{tab:vec}
\end{table*}

Assume that the seven particle species in this approach form a basis $B=\{K^{-}, \bar{p},\, \bar{\Lambda},\, \phi,\, \overline{\Xi}^{+}, \Omega^{-}, \overline{\Omega}^{+}\}$ of a 7-dimensional vector space, {\bf R}$^7$, where the elements of $B$ are called basis vectors in this space. Each combination or index in Table~\ref{tab:delq_dels} is a vector in {\bf R}$^7$, as shown in Table~\ref{tab:vec}, and all the indices together form a set of vectors, 
%% $\boldsymbol{V}=\{\boldsymbol{v_1, v_2,\ldots, v_r}\}$, 
$\boldsymbol{V}=\{\boldsymbol{v}_1, \boldsymbol{v}_2,\ldots, \boldsymbol{v}_r\}$,
where $r$ is the total number of vectors in the set.

The 10 indices in Table ~\ref{tab:delq_dels} are not linearly independent: 
(7) - (6) = (2), (3) + (5) = (6) and (10) - (8) = 1/3 (9). For each dependent relationship, one index should be removed. 
%% to check the linear dependence. 
Suppose that (2), (3) and (8) are removed. This is not a unique choice. The linear dependence of the remaining seven indices (1, 4, 5, 6, 7, 9 and 10) are checked using linear algebra. Therefore, the seven vectors from these seven indices form the set whose linear dependence ought to be evaluated. 

%From the theory of vector spaces in linear algebra, 
The linear dependence of a set of vectors 
$\boldsymbol{V}=\{\boldsymbol{v}_1, \boldsymbol{v}_2,\ldots, \boldsymbol{v}_r\}$,
requires that given a set of non-zero scalars
$(\alpha_1, \alpha_2, \ldots, \alpha_r)$ they can be written a as linear sum  
 \begin{equation}
      \sum_{i=1}^{r} \alpha_i \boldsymbol{v}_i = \bold{0} 
      \label{lindep}
 \end{equation}
 where $r$ is the total number of vectors in the set $\boldsymbol{V}$ (in our case $r=7$, for the seven vectors). Here $\bold{0}$ is a null vector in the same vector space. This implies that at least one vector is redundant and can be expressed as a linear combination of the others. However, if Eq.(\ref{lindep}) holds only for the case when all the coefficients are zero, the vectors $\boldsymbol{V}$ are said to be linearly independent~\cite{riley2006mathematical}. 
 %
 %This implies that at least one of the scalars is nonzero, say $a_1 \ne 0$. Then Eq.~(\ref{lindep}) leads to
 
% \begin{equation}
 %      \boldsymbol{v}_1 = \frac{-a_2}{a_1}\boldsymbol{v}_2 + \ldots + \frac{-a_r}{a_1}\boldsymbol{v}_r
  %    \label{lindep2}
% \end{equation}
%if $r>1$, while $\boldsymbol{v}_1=\bold{0}$ if $r=0$. Therefore, the set of vectors, $\boldsymbol{V}$, is linearly dependent if and only if one of them is zero or a linear combination of the others.

Each vector of $\boldsymbol{V}$ can be represented as a column matrix of dimension 7$\times$1, where 7 is the dimension of the vector space in the present case. Then Eq.~(\ref{lindep}) turns into a matrix equation where the seven vectors together form a matrix, $M$, of dimension 7$\times$7 and the scalars $\alpha_1, \alpha_2, \ldots, \alpha_7$ constitute a column matrix, $A$, with dimensions 7$\times$1, i.e.,

\begin{equation}
      MA=O,
\label{lindep3}
\end{equation}
with $O$ being a null matrix of dimensions 7$\times$1.

In order to evaluate the matrix equation, Eq.~(\ref{lindep3}), the matrix $M$ should be expressed in row-reduced echelon form by several row and column operations. Finally, Eq.~(\ref{lindep3}) with the row-reduced form of $M$ provides the values and relations among the scalars $\alpha_1, \alpha_2, \ldots, \alpha_7$. There is also an alternative method to evaluate the scalars which exploits determinants of the original matrix $M$.

Based on the aforementioned method of linear algebra, we find that the seven indices (1, 4, 5, 6, 7, 9 and 10) are not linearly independent. The number of vectors in the set $\boldsymbol{V}$ can be reduced repeatedly until an independent vector subset is identified. It is thus found that the five indices 4, 5, 6, 9 and 10 are linearly independent. Note that other sets of independent combinations exist; e.g., 1, 4, 5, 9 and 10; also 1, 5, 6, 9 and 10.

%Based on the aforementioned method of linear algebra, it is found that the seven indices (1, 4, 5, 6, 7, 9 and 10) are not linearly independent. If index 9 is omitted, the remaining indices (1, 4, 5, 6, 7, 10) are linearly dependent. In order to find an independent vector subset, the number of vectors in the set $\boldsymbol{V}$ can be reduced again. Indices 4 and 7 are removed this time, and a new vector set with indices 1, 5, 6, 9 and 10 is formed. Repeating the above exercise, it is found that the indices 1, 5, 6, 9 and 10 are linearly independent. At the end, we only have five (1, 5, 6, 9 and 10) linearly independent combinations among 10 combinations in Table~\ref{tab:delq_dels}. Note that other sets of independent combinations exist; e.g., other possible sets of independent combinations are: 1, 4, 5, 9 and 10; also 4, 5, 6, 9 and 10.

\bibliography{main}

\begin{thebibliography}{46}%
\makeatletter
\providecommand \@ifxundefined [1]{%
 \@ifx{#1\undefined}
}%
\providecommand \@ifnum [1]{%
 \ifnum #1\expandafter \@firstoftwo
 \else \expandafter \@secondoftwo
 \fi
}%
\providecommand \@ifx [1]{%
 \ifx #1\expandafter \@firstoftwo
 \else \expandafter \@secondoftwo
 \fi
}%
\providecommand \natexlab [1]{#1}%
\providecommand \enquote  [1]{``#1''}%
\providecommand \bibnamefont  [1]{#1}%
\providecommand \bibfnamefont [1]{#1}%
\providecommand \citenamefont [1]{#1}%
\providecommand \href@noop [0]{\@secondoftwo}%
\providecommand \href [0]{\begingroup \@sanitize@url \@href}%
\providecommand \@href[1]{\@@startlink{#1}\@@href}%
\providecommand \@@href[1]{\endgroup#1\@@endlink}%
\providecommand \@sanitize@url [0]{\catcode `\\12\catcode `\$12\catcode
  `\&12\catcode `\#12\catcode `\^12\catcode `\_12\catcode `\%12\relax}%
\providecommand \@@startlink[1]{}%
\providecommand \@@endlink[0]{}%
\providecommand \url  [0]{\begingroup\@sanitize@url \@url }%
\providecommand \@url [1]{\endgroup\@href {#1}{\urlprefix }}%
\providecommand \urlprefix  [0]{URL }%
\providecommand \Eprint [0]{\href }%
\providecommand \doibase [0]{http://dx.doi.org/}%
\providecommand \selectlanguage [0]{\@gobble}%
\providecommand \bibinfo  [0]{\@secondoftwo}%
\providecommand \bibfield  [0]{\@secondoftwo}%
\providecommand \translation [1]{[#1]}%
\providecommand \BibitemOpen [0]{}%
\providecommand \bibitemStop [0]{}%
\providecommand \bibitemNoStop [0]{.\EOS\space}%
\providecommand \EOS [0]{\spacefactor3000\relax}%
\providecommand \BibitemShut  [1]{\csname bibitem#1\endcsname}%
\let\auto@bib@innerbib\@empty
%</preamble>
\bibitem [{\citenamefont {Voloshin}\ and\ \citenamefont
  {Zhang}(1996)}]{Voloshin:1994mz}%
  \BibitemOpen
  \bibfield  {author} {\bibinfo {author} {\bibfnamefont {S.}~\bibnamefont
  {Voloshin}}\ and\ \bibinfo {author} {\bibfnamefont {Y.}~\bibnamefont
  {Zhang}},\ }\bibfield  {title} {\enquote {\bibinfo {title} {{Flow study in
  relativistic nuclear collisions by Fourier expansion of Azimuthal particle
  distributions}},}\ }\href {\doibase 10.1007/s002880050141} {\bibfield
  {journal} {\bibinfo  {journal} {Z. Phys. C}\ }\textbf {\bibinfo {volume}
  {70}},\ \bibinfo {pages} {665--672} (\bibinfo {year} {1996})},\ \Eprint
  {http://arxiv.org/abs/hep-ph/9407282} {arXiv:hep-ph/9407282} \BibitemShut
  {NoStop}%
\bibitem [{\citenamefont {Poskanzer}\ and\ \citenamefont
  {Voloshin}(1998)}]{Poskanzer:1998yz}%
  \BibitemOpen
  \bibfield  {author} {\bibinfo {author} {\bibfnamefont {Arthur~M.}\
  \bibnamefont {Poskanzer}}\ and\ \bibinfo {author} {\bibfnamefont {S.~A.}\
  \bibnamefont {Voloshin}},\ }\bibfield  {title} {\enquote {\bibinfo {title}
  {{Methods for analyzing anisotropic flow in relativistic nuclear
  collisions}},}\ }\href {\doibase 10.1103/PhysRevC.58.1671} {\bibfield
  {journal} {\bibinfo  {journal} {Phys. Rev. C}\ }\textbf {\bibinfo {volume}
  {58}},\ \bibinfo {pages} {1671--1678} (\bibinfo {year} {1998})},\ \Eprint
  {http://arxiv.org/abs/nucl-ex/9805001} {arXiv:nucl-ex/9805001} \BibitemShut
  {NoStop}%
\bibitem [{\citenamefont {G\"ursoy}\ \emph {et~al.}(2018)\citenamefont
  {G\"ursoy}, \citenamefont {Kharzeev}, \citenamefont {Marcus}, \citenamefont
  {Rajagopal},\ and\ \citenamefont {Shen}}]{Gursoy:2018yai}%
  \BibitemOpen
  \bibfield  {author} {\bibinfo {author} {\bibfnamefont {Umut}\ \bibnamefont
  {G\"ursoy}}, \bibinfo {author} {\bibfnamefont {Dmitri}\ \bibnamefont
  {Kharzeev}}, \bibinfo {author} {\bibfnamefont {Eric}\ \bibnamefont {Marcus}},
  \bibinfo {author} {\bibfnamefont {Krishna}\ \bibnamefont {Rajagopal}}, \ and\
  \bibinfo {author} {\bibfnamefont {Chun}\ \bibnamefont {Shen}},\ }\bibfield
  {title} {\enquote {\bibinfo {title} {{Charge-dependent Flow Induced by
  Magnetic and Electric Fields in Heavy Ion Collisions}},}\ }\href {\doibase
  10.1103/PhysRevC.98.055201} {\bibfield  {journal} {\bibinfo  {journal} {Phys.
  Rev. C}\ }\textbf {\bibinfo {volume} {98}},\ \bibinfo {pages} {055201}
  (\bibinfo {year} {2018})},\ \Eprint {http://arxiv.org/abs/1806.05288}
  {arXiv:1806.05288 [hep-ph]} \BibitemShut {NoStop}%
\bibitem [{\citenamefont {G{\"u}rsoy}\ \emph {et~al.}(2014)\citenamefont
  {G{\"u}rsoy}, \citenamefont {Kharzeev},\ and\ \citenamefont
  {Rajagopal}}]{Gursoy:2014aka}%
  \BibitemOpen
  \bibfield  {author} {\bibinfo {author} {\bibfnamefont {Umut}\ \bibnamefont
  {G{\"u}rsoy}}, \bibinfo {author} {\bibfnamefont {Dmitri}\ \bibnamefont
  {Kharzeev}}, \ and\ \bibinfo {author} {\bibfnamefont {Krishna}\ \bibnamefont
  {Rajagopal}},\ }\bibfield  {title} {\enquote {\bibinfo {title}
  {{Magnetohydrodynamics, charged currents and directed flow in heavy ion
  collisions}},}\ }\href {\doibase 10.1103/PhysRevC.89.054905} {\bibfield
  {journal} {\bibinfo  {journal} {Phys. Rev. C}\ }\textbf {\bibinfo {volume}
  {89}},\ \bibinfo {pages} {054905} (\bibinfo {year} {2014})},\ \Eprint
  {http://arxiv.org/abs/1401.3805} {arXiv:1401.3805 [hep-ph]} \BibitemShut
  {NoStop}%
\bibitem [{\citenamefont {Dubla}\ \emph {et~al.}(2020)\citenamefont {Dubla},
  \citenamefont {G\"ursoy},\ and\ \citenamefont {Snellings}}]{Dubla:2020bdz}%
  \BibitemOpen
  \bibfield  {author} {\bibinfo {author} {\bibfnamefont {Andrea}\ \bibnamefont
  {Dubla}}, \bibinfo {author} {\bibfnamefont {Umut}\ \bibnamefont {G\"ursoy}},
  \ and\ \bibinfo {author} {\bibfnamefont {Raimond}\ \bibnamefont
  {Snellings}},\ }\bibfield  {title} {\enquote {\bibinfo {title}
  {{Charge-dependent flow as evidence of strong electromagnetic fields in
  heavy-ion collisions}},}\ }\href {\doibase 10.1142/S0217732320503241}
  {\bibfield  {journal} {\bibinfo  {journal} {Mod. Phys. Lett. A}\ }\textbf
  {\bibinfo {volume} {35}},\ \bibinfo {pages} {2050324} (\bibinfo {year}
  {2020})},\ \Eprint {http://arxiv.org/abs/2009.09727} {arXiv:2009.09727
  [hep-ph]} \BibitemShut {NoStop}%
\bibitem [{\citenamefont {Voronyuk}\ \emph {et~al.}(2014)\citenamefont
  {Voronyuk}, \citenamefont {Toneev}, \citenamefont {Voloshin},\ and\
  \citenamefont {Cassing}}]{Voronyuk:2014rna}%
  \BibitemOpen
  \bibfield  {author} {\bibinfo {author} {\bibfnamefont {V.}~\bibnamefont
  {Voronyuk}}, \bibinfo {author} {\bibfnamefont {V.~D.}\ \bibnamefont
  {Toneev}}, \bibinfo {author} {\bibfnamefont {S.~A.}\ \bibnamefont
  {Voloshin}}, \ and\ \bibinfo {author} {\bibfnamefont {W.}~\bibnamefont
  {Cassing}},\ }\bibfield  {title} {\enquote {\bibinfo {title}
  {{Charge-dependent directed flow in asymmetric nuclear collisions}},}\ }\href
  {\doibase 10.1103/PhysRevC.90.064903} {\bibfield  {journal} {\bibinfo
  {journal} {Phys. Rev. C}\ }\textbf {\bibinfo {volume} {90}},\ \bibinfo
  {pages} {064903} (\bibinfo {year} {2014})},\ \Eprint
  {http://arxiv.org/abs/1410.1402} {arXiv:1410.1402 [nucl-th]} \BibitemShut
  {NoStop}%
\bibitem [{\citenamefont {Toneev}\ \emph {et~al.}(2017)\citenamefont {Toneev},
  \citenamefont {Voronyuk}, \citenamefont {Kolomeitsev},\ and\ \citenamefont
  {Cassing}}]{Toneev:2016bri}%
  \BibitemOpen
  \bibfield  {author} {\bibinfo {author} {\bibfnamefont {V.~D.}\ \bibnamefont
  {Toneev}}, \bibinfo {author} {\bibfnamefont {V.}~\bibnamefont {Voronyuk}},
  \bibinfo {author} {\bibfnamefont {E.~E.}\ \bibnamefont {Kolomeitsev}}, \ and\
  \bibinfo {author} {\bibfnamefont {W.}~\bibnamefont {Cassing}},\ }\bibfield
  {title} {\enquote {\bibinfo {title} {{Directed flow in asymmetric
  nucleus-nucleus collisions and the inverse Landau-Pomeranchuk-Migdal
  effect}},}\ }\href {\doibase 10.1103/PhysRevC.95.034911} {\bibfield
  {journal} {\bibinfo  {journal} {Phys. Rev. C}\ }\textbf {\bibinfo {volume}
  {95}},\ \bibinfo {pages} {034911} (\bibinfo {year} {2017})},\ \Eprint
  {http://arxiv.org/abs/1610.06319} {arXiv:1610.06319 [nucl-th]} \BibitemShut
  {NoStop}%
\bibitem [{\citenamefont {Oliva}\ \emph {et~al.}(2020)\citenamefont {Oliva},
  \citenamefont {Moreau}, \citenamefont {Voronyuk},\ and\ \citenamefont
  {Bratkovskaya}}]{Oliva:2019kin}%
  \BibitemOpen
  \bibfield  {author} {\bibinfo {author} {\bibfnamefont {Lucia}\ \bibnamefont
  {Oliva}}, \bibinfo {author} {\bibfnamefont {Pierre}\ \bibnamefont {Moreau}},
  \bibinfo {author} {\bibfnamefont {Vadim}\ \bibnamefont {Voronyuk}}, \ and\
  \bibinfo {author} {\bibfnamefont {Elena}\ \bibnamefont {Bratkovskaya}},\
  }\bibfield  {title} {\enquote {\bibinfo {title} {{Influence of
  electromagnetic fields in proton-nucleus collisions at relativistic
  energy}},}\ }\href {\doibase 10.1103/PhysRevC.101.014917} {\bibfield
  {journal} {\bibinfo  {journal} {Phys. Rev. C}\ }\textbf {\bibinfo {volume}
  {101}},\ \bibinfo {pages} {014917} (\bibinfo {year} {2020})},\ \Eprint
  {http://arxiv.org/abs/1909.06770} {arXiv:1909.06770 [nucl-th]} \BibitemShut
  {NoStop}%
\bibitem [{\citenamefont {Kharzeev}\ \emph {et~al.}(2008)\citenamefont
  {Kharzeev}, \citenamefont {McLerran},\ and\ \citenamefont
  {Warringa}}]{Kharzeev:2007jp}%
  \BibitemOpen
  \bibfield  {author} {\bibinfo {author} {\bibfnamefont {Dmitri~E.}\
  \bibnamefont {Kharzeev}}, \bibinfo {author} {\bibfnamefont {Larry~D.}\
  \bibnamefont {McLerran}}, \ and\ \bibinfo {author} {\bibfnamefont
  {Harmen~J.}\ \bibnamefont {Warringa}},\ }\bibfield  {title} {\enquote
  {\bibinfo {title} {{The Effects of topological charge change in heavy ion
  collisions: 'Event by event P and CP violation'}},}\ }\href {\doibase
  10.1016/j.nuclphysa.2008.02.298} {\bibfield  {journal} {\bibinfo  {journal}
  {Nucl. Phys.}\ }\textbf {\bibinfo {volume} {A803}},\ \bibinfo {pages}
  {227--253} (\bibinfo {year} {2008})},\ \Eprint
  {http://arxiv.org/abs/0711.0950} {arXiv:0711.0950 [hep-ph]} \BibitemShut
  {NoStop}%
%%CITATION = ARXIV:0711.0950;%%
\bibitem [{\citenamefont {Skokov}\ \emph {et~al.}(2009)\citenamefont {Skokov},
  \citenamefont {Illarionov},\ and\ \citenamefont {Toneev}}]{Skokov:2009qp}%
  \BibitemOpen
  \bibfield  {author} {\bibinfo {author} {\bibfnamefont {V.}~\bibnamefont
  {Skokov}}, \bibinfo {author} {\bibfnamefont {A.~{\relax Yu}.}\ \bibnamefont
  {Illarionov}}, \ and\ \bibinfo {author} {\bibfnamefont {V.}~\bibnamefont
  {Toneev}},\ }\bibfield  {title} {\enquote {\bibinfo {title} {{Estimate of the
  magnetic field strength in heavy-ion collisions}},}\ }\href {\doibase
  10.1142/S0217751X09047570} {\bibfield  {journal} {\bibinfo  {journal} {Int.
  J. Mod. Phys.}\ }\textbf {\bibinfo {volume} {A24}},\ \bibinfo {pages}
  {5925--5932} (\bibinfo {year} {2009})},\ \Eprint
  {http://arxiv.org/abs/0907.1396} {arXiv:0907.1396 [nucl-th]} \BibitemShut
  {NoStop}%
%%CITATION = ARXIV:0907.1396;%%
\bibitem [{\citenamefont {Bzdak}\ and\ \citenamefont
  {Skokov}(2012)}]{Bzdak:2011yy}%
  \BibitemOpen
  \bibfield  {author} {\bibinfo {author} {\bibfnamefont {Adam}\ \bibnamefont
  {Bzdak}}\ and\ \bibinfo {author} {\bibfnamefont {Vladimir}\ \bibnamefont
  {Skokov}},\ }\bibfield  {title} {\enquote {\bibinfo {title} {{Event-by-event
  fluctuations of magnetic and electric fields in heavy ion collisions}},}\
  }\href {\doibase 10.1016/j.physletb.2012.02.065} {\bibfield  {journal}
  {\bibinfo  {journal} {Phys.Lett.}\ }\textbf {\bibinfo {volume} {B710}},\
  \bibinfo {pages} {171--174} (\bibinfo {year} {2012})},\ \Eprint
  {http://arxiv.org/abs/1111.1949} {arXiv:1111.1949 [hep-ph]} \BibitemShut
  {NoStop}%
%%CITATION = ARXIV:1111.1949;%%
\bibitem [{\citenamefont {Deng}\ and\ \citenamefont
  {Huang}(2012)}]{Deng:2012pc}%
  \BibitemOpen
  \bibfield  {author} {\bibinfo {author} {\bibfnamefont {Wei-Tian}\
  \bibnamefont {Deng}}\ and\ \bibinfo {author} {\bibfnamefont {Xu-Guang}\
  \bibnamefont {Huang}},\ }\bibfield  {title} {\enquote {\bibinfo {title}
  {{Event-by-event generation of electromagnetic fields in heavy-ion
  collisions}},}\ }\href {\doibase 10.1103/PhysRevC.85.044907} {\bibfield
  {journal} {\bibinfo  {journal} {Phys.Rev.}\ }\textbf {\bibinfo {volume}
  {C85}},\ \bibinfo {pages} {044907} (\bibinfo {year} {2012})},\ \Eprint
  {http://arxiv.org/abs/1201.5108} {arXiv:1201.5108 [nucl-th]} \BibitemShut
  {NoStop}%
%%CITATION = ARXIV:1201.5108;%%
\bibitem [{\citenamefont {Bloczynski}\ \emph {et~al.}(2013)\citenamefont
  {Bloczynski}, \citenamefont {Huang}, \citenamefont {Zhang},\ and\
  \citenamefont {Liao}}]{Bloczynski:2012en}%
  \BibitemOpen
  \bibfield  {author} {\bibinfo {author} {\bibfnamefont {John}\ \bibnamefont
  {Bloczynski}}, \bibinfo {author} {\bibfnamefont {Xu-Guang}\ \bibnamefont
  {Huang}}, \bibinfo {author} {\bibfnamefont {Xilin}\ \bibnamefont {Zhang}}, \
  and\ \bibinfo {author} {\bibfnamefont {Jinfeng}\ \bibnamefont {Liao}},\
  }\bibfield  {title} {\enquote {\bibinfo {title} {{Azimuthally fluctuating
  magnetic field and its impacts on observables in heavy-ion collisions}},}\
  }\href {\doibase 10.1016/j.physletb.2012.12.030} {\bibfield  {journal}
  {\bibinfo  {journal} {Phys. Lett. B}\ }\textbf {\bibinfo {volume} {718}},\
  \bibinfo {pages} {1529--1535} (\bibinfo {year} {2013})},\ \Eprint
  {http://arxiv.org/abs/1209.6594} {arXiv:1209.6594 [nucl-th]} \BibitemShut
  {NoStop}%
\bibitem [{\citenamefont {Tuchin}(2013)}]{Tuchin:2013ie}%
  \BibitemOpen
  \bibfield  {author} {\bibinfo {author} {\bibfnamefont {Kirill}\ \bibnamefont
  {Tuchin}},\ }\bibfield  {title} {\enquote {\bibinfo {title} {{Particle
  production in strong electromagnetic fields in relativistic heavy-ion
  collisions}},}\ }\href {\doibase 10.1155/2013/490495} {\bibfield  {journal}
  {\bibinfo  {journal} {Adv. High Energy Phys.}\ }\textbf {\bibinfo {volume}
  {2013}},\ \bibinfo {pages} {490495} (\bibinfo {year} {2013})},\ \Eprint
  {http://arxiv.org/abs/1301.0099} {arXiv:1301.0099 [hep-ph]} \BibitemShut
  {NoStop}%
\bibitem [{\citenamefont {Bloczynski}\ \emph {et~al.}(2015)\citenamefont
  {Bloczynski}, \citenamefont {Huang}, \citenamefont {Zhang},\ and\
  \citenamefont {Liao}}]{Bloczynski:2013mca}%
  \BibitemOpen
  \bibfield  {author} {\bibinfo {author} {\bibfnamefont {John}\ \bibnamefont
  {Bloczynski}}, \bibinfo {author} {\bibfnamefont {Xu-Guang}\ \bibnamefont
  {Huang}}, \bibinfo {author} {\bibfnamefont {Xilin}\ \bibnamefont {Zhang}}, \
  and\ \bibinfo {author} {\bibfnamefont {Jinfeng}\ \bibnamefont {Liao}},\
  }\bibfield  {title} {\enquote {\bibinfo {title} {{Charge-dependent azimuthal
  correlations from AuAu to UU collisions}},}\ }\href {\doibase
  10.1016/j.nuclphysa.2015.03.012} {\bibfield  {journal} {\bibinfo  {journal}
  {Nucl. Phys. A}\ }\textbf {\bibinfo {volume} {939}},\ \bibinfo {pages}
  {85--100} (\bibinfo {year} {2015})},\ \Eprint
  {http://arxiv.org/abs/1311.5451} {arXiv:1311.5451 [nucl-th]} \BibitemShut
  {NoStop}%
\bibitem [{\citenamefont {McLerran}\ and\ \citenamefont
  {Skokov}(2014)}]{McLerran:2013hla}%
  \BibitemOpen
  \bibfield  {author} {\bibinfo {author} {\bibfnamefont {L.}~\bibnamefont
  {McLerran}}\ and\ \bibinfo {author} {\bibfnamefont {V.}~\bibnamefont
  {Skokov}},\ }\bibfield  {title} {\enquote {\bibinfo {title} {{Comments About
  the Electromagnetic Field in Heavy-Ion Collisions}},}\ }\href {\doibase
  10.1016/j.nuclphysa.2014.05.008} {\bibfield  {journal} {\bibinfo  {journal}
  {Nucl. Phys. A}\ }\textbf {\bibinfo {volume} {929}},\ \bibinfo {pages}
  {184--190} (\bibinfo {year} {2014})},\ \Eprint
  {http://arxiv.org/abs/1305.0774} {arXiv:1305.0774 [hep-ph]} \BibitemShut
  {NoStop}%
\bibitem [{\citenamefont {Sun}\ \emph {et~al.}(2019)\citenamefont {Sun},
  \citenamefont {Wang}, \citenamefont {Li},\ and\ \citenamefont
  {Wang}}]{Sun:2019hao}%
  \BibitemOpen
  \bibfield  {author} {\bibinfo {author} {\bibfnamefont {Yuliang}\ \bibnamefont
  {Sun}}, \bibinfo {author} {\bibfnamefont {Yongjia}\ \bibnamefont {Wang}},
  \bibinfo {author} {\bibfnamefont {Qingfeng}\ \bibnamefont {Li}}, \ and\
  \bibinfo {author} {\bibfnamefont {Fuqiang}\ \bibnamefont {Wang}},\ }\bibfield
   {title} {\enquote {\bibinfo {title} {{Effect of internal magnetic field on
  collective flow in heavy ion collisions at intermediate energies}},}\ }\href
  {\doibase 10.1103/PhysRevC.99.064607} {\bibfield  {journal} {\bibinfo
  {journal} {Phys. Rev. C}\ }\textbf {\bibinfo {volume} {99}},\ \bibinfo
  {pages} {064607} (\bibinfo {year} {2019})},\ \Eprint
  {http://arxiv.org/abs/1905.12492} {arXiv:1905.12492 [nucl-th]} \BibitemShut
  {NoStop}%
\bibitem [{\citenamefont {Das}\ \emph {et~al.}(2017)\citenamefont {Das},
  \citenamefont {Plumari}, \citenamefont {Chatterjee}, \citenamefont {Alam},
  \citenamefont {Scardina},\ and\ \citenamefont {Greco}}]{Das:2016cwd}%
  \BibitemOpen
  \bibfield  {author} {\bibinfo {author} {\bibfnamefont {Santosh~K.}\
  \bibnamefont {Das}}, \bibinfo {author} {\bibfnamefont {Salvatore}\
  \bibnamefont {Plumari}}, \bibinfo {author} {\bibfnamefont {Sandeep}\
  \bibnamefont {Chatterjee}}, \bibinfo {author} {\bibfnamefont {Jane}\
  \bibnamefont {Alam}}, \bibinfo {author} {\bibfnamefont {Francesco}\
  \bibnamefont {Scardina}}, \ and\ \bibinfo {author} {\bibfnamefont {Vincenzo}\
  \bibnamefont {Greco}},\ }\bibfield  {title} {\enquote {\bibinfo {title}
  {{Directed Flow of Charm Quarks as a Witness of the Initial Strong Magnetic
  Field in Ultra-Relativistic Heavy Ion Collisions}},}\ }\href {\doibase
  10.1016/j.physletb.2017.02.046} {\bibfield  {journal} {\bibinfo  {journal}
  {Phys. Lett. B}\ }\textbf {\bibinfo {volume} {768}},\ \bibinfo {pages}
  {260--264} (\bibinfo {year} {2017})},\ \Eprint
  {http://arxiv.org/abs/1608.02231} {arXiv:1608.02231 [nucl-th]} \BibitemShut
  {NoStop}%
\bibitem [{\citenamefont {Adam}\ \emph {et~al.}(2019)\citenamefont {Adam} \emph
  {et~al.}}]{Adam:2019wnk}%
  \BibitemOpen
  \bibfield  {author} {\bibinfo {author} {\bibfnamefont {Jaroslav}\
  \bibnamefont {Adam}} \emph {et~al.} (\bibinfo {collaboration} {STAR
  Collaboration}),\ }\bibfield  {title} {\enquote {\bibinfo {title} {{First
  Observation of the Directed Flow of $D^{0}$ and $\overline{D^0}$ in Au+Au
  Collisions at $\sqrt{s_{\rm NN}}$ = 200 GeV}},}\ }\href {\doibase
  10.1103/PhysRevLett.123.162301} {\bibfield  {journal} {\bibinfo  {journal}
  {Phys. Rev. Lett.}\ }\textbf {\bibinfo {volume} {123}},\ \bibinfo {pages}
  {162301} (\bibinfo {year} {2019})},\ \Eprint
  {http://arxiv.org/abs/1905.02052} {arXiv:1905.02052 [nucl-ex]} \BibitemShut
  {NoStop}%
\bibitem [{\citenamefont {Acharya}\ \emph {et~al.}(2020)\citenamefont {Acharya}
  \emph {et~al.}}]{Acharya:2019ijj}%
  \BibitemOpen
  \bibfield  {author} {\bibinfo {author} {\bibfnamefont {Shreyasi}\
  \bibnamefont {Acharya}} \emph {et~al.} (\bibinfo {collaboration} {ALICE
  Collaboration}),\ }\bibfield  {title} {\enquote {\bibinfo {title} {{Probing
  the effects of strong electromagnetic fields with charge-dependent directed
  flow in Pb-Pb collisions at the LHC}},}\ }\href {\doibase
  10.1103/PhysRevLett.125.022301} {\bibfield  {journal} {\bibinfo  {journal}
  {Phys. Rev. Lett.}\ }\textbf {\bibinfo {volume} {125}},\ \bibinfo {pages}
  {022301} (\bibinfo {year} {2020})},\ \Eprint
  {http://arxiv.org/abs/1910.14406} {arXiv:1910.14406 [nucl-ex]} \BibitemShut
  {NoStop}%
\bibitem [{sta()}]{starpsn0600}%
  \BibitemOpen
  \href@noop {} {}\bibinfo {note} {The STAR Heavy Flavor Tracker - Conceptual
  Design Report
  \href{https://drupal.star.bnl.gov/STAR/starnotes/public/sn0600}{\url{https://drupal.star.bnl.gov/STAR/starnotes/public/sn0600}}}\BibitemShut
  {NoStop}%
\bibitem [{\citenamefont {Adamczyk}\ \emph {et~al.}(2017)\citenamefont
  {Adamczyk} \emph {et~al.}}]{Adamczyk:2016eux}%
  \BibitemOpen
  \bibfield  {author} {\bibinfo {author} {\bibfnamefont {L.}~\bibnamefont
  {Adamczyk}} \emph {et~al.} (\bibinfo {collaboration} {STAR Collaboration}),\
  }\bibfield  {title} {\enquote {\bibinfo {title} {{Charge-dependent directed
  flow in Cu+Au collisions at $\sqrt{s_{_{NN}}}$ = 200 GeV}},}\ }\href
  {\doibase 10.1103/PhysRevLett.118.012301} {\bibfield  {journal} {\bibinfo
  {journal} {Phys. Rev. Lett.}\ }\textbf {\bibinfo {volume} {118}},\ \bibinfo
  {pages} {012301} (\bibinfo {year} {2017})},\ \Eprint
  {http://arxiv.org/abs/1608.04100} {arXiv:1608.04100 [nucl-ex]} \BibitemShut
  {NoStop}%
\bibitem [{\citenamefont {Adamczyk}\ \emph {et~al.}(2014)\citenamefont
  {Adamczyk} \emph {et~al.}}]{Adamczyk:2014ipa}%
  \BibitemOpen
  \bibfield  {author} {\bibinfo {author} {\bibfnamefont {L.}~\bibnamefont
  {Adamczyk}} \emph {et~al.} (\bibinfo {collaboration} {STAR Collaboration}),\
  }\bibfield  {title} {\enquote {\bibinfo {title} {{Beam-Energy Dependence of
  the Directed Flow of Protons, Antiprotons, and Pions in Au+Au Collisions}},}\
  }\href {\doibase 10.1103/PhysRevLett.112.162301} {\bibfield  {journal}
  {\bibinfo  {journal} {Phys. Rev. Lett.}\ }\textbf {\bibinfo {volume} {112}},\
  \bibinfo {pages} {162301} (\bibinfo {year} {2014})},\ \Eprint
  {http://arxiv.org/abs/1401.3043} {arXiv:1401.3043 [nucl-ex]} \BibitemShut
  {NoStop}%
\bibitem [{\citenamefont {Adamczyk}\ \emph {et~al.}(2018)\citenamefont
  {Adamczyk} \emph {et~al.}}]{Adamczyk:2017nxg}%
  \BibitemOpen
  \bibfield  {author} {\bibinfo {author} {\bibfnamefont {Leszek}\ \bibnamefont
  {Adamczyk}} \emph {et~al.} (\bibinfo {collaboration} {STAR Collaboration}),\
  }\bibfield  {title} {\enquote {\bibinfo {title} {{Beam-Energy Dependence of
  Directed Flow of $\Lambda$, $\bar{\Lambda}$, $K^\pm$, $K^0_s$ and $\phi$ in
  Au+Au Collisions}},}\ }\href {\doibase 10.1103/PhysRevLett.120.062301}
  {\bibfield  {journal} {\bibinfo  {journal} {Phys. Rev. Lett.}\ }\textbf
  {\bibinfo {volume} {120}},\ \bibinfo {pages} {062301} (\bibinfo {year}
  {2018})},\ \Eprint {http://arxiv.org/abs/1708.07132} {arXiv:1708.07132
  [hep-ex]} \BibitemShut {NoStop}%
\bibitem [{\citenamefont {Dunlop}\ \emph {et~al.}(2011)\citenamefont {Dunlop},
  \citenamefont {Lisa},\ and\ \citenamefont {Sorensen}}]{Dunlop:2011cf}%
  \BibitemOpen
  \bibfield  {author} {\bibinfo {author} {\bibfnamefont {J.~C.}\ \bibnamefont
  {Dunlop}}, \bibinfo {author} {\bibfnamefont {M.~A.}\ \bibnamefont {Lisa}}, \
  and\ \bibinfo {author} {\bibfnamefont {P.}~\bibnamefont {Sorensen}},\
  }\bibfield  {title} {\enquote {\bibinfo {title} {{Constituent quark scaling
  violation due to baryon number transport}},}\ }\href {\doibase
  10.1103/PhysRevC.84.044914} {\bibfield  {journal} {\bibinfo  {journal} {Phys.
  Rev. C}\ }\textbf {\bibinfo {volume} {84}},\ \bibinfo {pages} {044914}
  (\bibinfo {year} {2011})},\ \Eprint {http://arxiv.org/abs/1107.3078}
  {arXiv:1107.3078 [hep-ph]} \BibitemShut {NoStop}%
\bibitem [{\citenamefont {Guo}\ \emph {et~al.}(2012)\citenamefont {Guo},
  \citenamefont {Liu},\ and\ \citenamefont {Tang}}]{Guo:2012qi}%
  \BibitemOpen
  \bibfield  {author} {\bibinfo {author} {\bibfnamefont {Yao}\ \bibnamefont
  {Guo}}, \bibinfo {author} {\bibfnamefont {Feng}\ \bibnamefont {Liu}}, \ and\
  \bibinfo {author} {\bibfnamefont {Aihong}\ \bibnamefont {Tang}},\ }\bibfield
  {title} {\enquote {\bibinfo {title} {{Directed flow of transported and
  non-transported protons in Au+Au collisions from UrQMD model}},}\ }\href
  {\doibase 10.1103/PhysRevC.86.044901} {\bibfield  {journal} {\bibinfo
  {journal} {Phys. Rev. C}\ }\textbf {\bibinfo {volume} {86}},\ \bibinfo
  {pages} {044901} (\bibinfo {year} {2012})},\ \Eprint
  {http://arxiv.org/abs/1206.2246} {arXiv:1206.2246 [nucl-ex]} \BibitemShut
  {NoStop}%
\bibitem [{\citenamefont {Wang}(2019)}]{Wang:2018pqx}%
  \BibitemOpen
  \bibfield  {author} {\bibinfo {author} {\bibfnamefont {Gang}\ \bibnamefont
  {Wang}} (\bibinfo {collaboration} {STAR Collaboration}),\ }\bibfield  {title}
  {\enquote {\bibinfo {title} {{Directed flow of quarks from the RHIC Beam
  Energy Scan measured by STAR}},}\ }\href {\doibase
  10.1016/j.nuclphysa.2018.08.036} {\bibfield  {journal} {\bibinfo  {journal}
  {Nucl. Phys. A}\ }\textbf {\bibinfo {volume} {982}},\ \bibinfo {pages}
  {415--418} (\bibinfo {year} {2019})},\ \Eprint
  {http://arxiv.org/abs/1807.05565} {arXiv:1807.05565 [nucl-ex]} \BibitemShut
  {NoStop}%
\bibitem [{\citenamefont {Voloshin}\ and\ \citenamefont
  {Niida}(2016)}]{Voloshin:2016ppr}%
  \BibitemOpen
  \bibfield  {author} {\bibinfo {author} {\bibfnamefont {Sergei~A.}\
  \bibnamefont {Voloshin}}\ and\ \bibinfo {author} {\bibfnamefont {Takafumi}\
  \bibnamefont {Niida}},\ }\bibfield  {title} {\enquote {\bibinfo {title}
  {{Ultrarelativistic nuclear collisions: Direction of spectator flow}},}\
  }\href {\doibase 10.1103/PhysRevC.94.021901} {\bibfield  {journal} {\bibinfo
  {journal} {Phys. Rev. C}\ }\textbf {\bibinfo {volume} {94}},\ \bibinfo
  {pages} {021901} (\bibinfo {year} {2016})},\ \Eprint
  {http://arxiv.org/abs/1604.04597} {arXiv:1604.04597 [nucl-th]} \BibitemShut
  {NoStop}%
\bibitem [{\citenamefont {Bratkovskaya}\ \emph {et~al.}(2011)\citenamefont
  {Bratkovskaya}, \citenamefont {Cassing}, \citenamefont {Konchakovski},\ and\
  \citenamefont {Linnyk}}]{Bratkovskaya2011PartonHadronStringDA}%
  \BibitemOpen
  \bibfield  {author} {\bibinfo {author} {\bibfnamefont {E.}~\bibnamefont
  {Bratkovskaya}}, \bibinfo {author} {\bibfnamefont {W.}~\bibnamefont
  {Cassing}}, \bibinfo {author} {\bibfnamefont {V.}~\bibnamefont
  {Konchakovski}}, \ and\ \bibinfo {author} {\bibfnamefont {O.}~\bibnamefont
  {Linnyk}},\ }\bibfield  {title} {\enquote {\bibinfo {title}
  {Parton-hadron-string dynamics at relativistic collider energies},}\
  }\href@noop {} {\bibfield  {journal} {\bibinfo  {journal} {Nucl. Phys. A}\
  }\textbf {\bibinfo {volume} {856}},\ \bibinfo {pages} {162--182} (\bibinfo
  {year} {2011})}\BibitemShut {NoStop}%
\bibitem [{\citenamefont {Konchakovski}\ \emph {et~al.}(2014)\citenamefont
  {Konchakovski}, \citenamefont {Cassing}, \citenamefont {Ivanov},\ and\
  \citenamefont {Toneev}}]{Konchakovski:2014gda}%
  \BibitemOpen
  \bibfield  {author} {\bibinfo {author} {\bibfnamefont {V.~P.}\ \bibnamefont
  {Konchakovski}}, \bibinfo {author} {\bibfnamefont {W.}~\bibnamefont
  {Cassing}}, \bibinfo {author} {\bibfnamefont {Yu.~B.}\ \bibnamefont
  {Ivanov}}, \ and\ \bibinfo {author} {\bibfnamefont {V.~D.}\ \bibnamefont
  {Toneev}},\ }\bibfield  {title} {\enquote {\bibinfo {title} {{Examination of
  the directed flow puzzle in heavy-ion collisions}},}\ }\href {\doibase
  10.1103/PhysRevC.90.014903} {\bibfield  {journal} {\bibinfo  {journal} {Phys.
  Rev. C}\ }\textbf {\bibinfo {volume} {90}},\ \bibinfo {pages} {014903}
  (\bibinfo {year} {2014})},\ \Eprint {http://arxiv.org/abs/1404.2765}
  {arXiv:1404.2765 [nucl-th]} \BibitemShut {NoStop}%
\bibitem [{\citenamefont {Palmese}\ and\ \citenamefont
  {Cassing}(2016)}]{Palmese:2016abq}%
  \BibitemOpen
  \bibfield  {author} {\bibinfo {author} {\bibfnamefont {A.}~\bibnamefont
  {Palmese}}\ and\ \bibinfo {author} {\bibfnamefont {W.}~\bibnamefont
  {Cassing}},\ }\bibfield  {title} {\enquote {\bibinfo {title} {{Directed flow
  in heavy-ion collisions from the PHSD transport approach}},}\ }\href
  {\doibase 10.1088/1742-6596/668/1/012075} {\bibfield  {journal} {\bibinfo
  {journal} {J. Phys. Conf. Ser.}\ }\textbf {\bibinfo {volume} {668}},\
  \bibinfo {pages} {012075} (\bibinfo {year} {2016})}\BibitemShut {NoStop}%
\bibitem [{\citenamefont {Voronyuk}\ \emph {et~al.}(2011)\citenamefont
  {Voronyuk}, \citenamefont {Toneev}, \citenamefont {Cassing}, \citenamefont
  {Bratkovskaya}, \citenamefont {Konchakovski},\ and\ \citenamefont
  {Voloshin}}]{Voronyuk:2011jd}%
  \BibitemOpen
  \bibfield  {author} {\bibinfo {author} {\bibfnamefont {V.}~\bibnamefont
  {Voronyuk}}, \bibinfo {author} {\bibfnamefont {V.~D.}\ \bibnamefont
  {Toneev}}, \bibinfo {author} {\bibfnamefont {W.}~\bibnamefont {Cassing}},
  \bibinfo {author} {\bibfnamefont {E.~L.}\ \bibnamefont {Bratkovskaya}},
  \bibinfo {author} {\bibfnamefont {V.~P.}\ \bibnamefont {Konchakovski}}, \
  and\ \bibinfo {author} {\bibfnamefont {S.~A.}\ \bibnamefont {Voloshin}},\
  }\bibfield  {title} {\enquote {\bibinfo {title} {{(Electro-)Magnetic field
  evolution in relativistic heavy-ion collisions}},}\ }\href {\doibase
  10.1103/PhysRevC.83.054911} {\bibfield  {journal} {\bibinfo  {journal} {Phys.
  Rev. C}\ }\textbf {\bibinfo {volume} {83}},\ \bibinfo {pages} {054911}
  (\bibinfo {year} {2011})},\ \Eprint {http://arxiv.org/abs/1103.4239}
  {arXiv:1103.4239 [nucl-th]} \BibitemShut {NoStop}%
\bibitem [{\citenamefont {Bratkovskaya}\ \emph {et~al.}()\citenamefont
  {Bratkovskaya}, \citenamefont {Oliva},\ and\ \citenamefont
  {Soloveva}}]{PHSDprivate}%
  \BibitemOpen
  \bibfield  {author} {\bibinfo {author} {\bibfnamefont {E.}~\bibnamefont
  {Bratkovskaya}}, \bibinfo {author} {\bibfnamefont {L.}~\bibnamefont {Oliva}},
  \ and\ \bibinfo {author} {\bibfnamefont {O.}~\bibnamefont {Soloveva}},\
  }\href@noop {} {}\bibinfo {howpublished} {Private Communication}\BibitemShut
  {NoStop}%
\bibitem [{\citenamefont {Soloveva}\ \emph {et~al.}(2020)\citenamefont
  {Soloveva}, \citenamefont {Moreau}, \citenamefont {Oliva}, \citenamefont
  {Voronyuk}, \citenamefont {Kireyeu}, \citenamefont {Song},\ and\
  \citenamefont {Bratkovskaya}}]{Soloveva:2020ozg}%
  \BibitemOpen
  \bibfield  {author} {\bibinfo {author} {\bibfnamefont {O.}~\bibnamefont
  {Soloveva}}, \bibinfo {author} {\bibfnamefont {P.}~\bibnamefont {Moreau}},
  \bibinfo {author} {\bibfnamefont {L.}~\bibnamefont {Oliva}}, \bibinfo
  {author} {\bibfnamefont {V.}~\bibnamefont {Voronyuk}}, \bibinfo {author}
  {\bibfnamefont {V.}~\bibnamefont {Kireyeu}}, \bibinfo {author} {\bibfnamefont
  {T.}~\bibnamefont {Song}}, \ and\ \bibinfo {author} {\bibfnamefont
  {E.}~\bibnamefont {Bratkovskaya}},\ }\bibfield  {title} {\enquote {\bibinfo
  {title} {{Exploring the partonic phase at finite chemical potential in and
  out-of equilibrium}},}\ }\href {\doibase 10.3390/particles3010015} {\bibfield
   {journal} {\bibinfo  {journal} {Particles}\ }\textbf {\bibinfo {volume}
  {3}},\ \bibinfo {pages} {178--192} (\bibinfo {year} {2020})},\ \Eprint
  {http://arxiv.org/abs/2001.05395} {arXiv:2001.05395 [nucl-th]} \BibitemShut
  {NoStop}%
\bibitem [{Sta()}]{StarBur:2021}%
  \BibitemOpen
  \href@noop {} {\enquote {\bibinfo {title} {{The STAR Beam Use Request For Run
  21}},}\ }\bibinfo {howpublished}
  {\url{https://drupal.star.bnl.gov/STAR/files/BUR2020_final.pdf}},\ \bibinfo
  {note} {accessed: 2021-09-21}\BibitemShut {NoStop}%
\bibitem [{\citenamefont {Adams}\ \emph {et~al.}(2020)\citenamefont {Adams}
  \emph {et~al.}}]{Adams:2019fpo}%
  \BibitemOpen
  \bibfield  {author} {\bibinfo {author} {\bibfnamefont {Joseph}\ \bibnamefont
  {Adams}} \emph {et~al.},\ }\bibfield  {title} {\enquote {\bibinfo {title}
  {{The STAR Event Plane Detector}},}\ }\href {\doibase
  10.1016/j.nima.2020.163970} {\bibfield  {journal} {\bibinfo  {journal} {Nucl.
  Instrum. Meth. A}\ }\textbf {\bibinfo {volume} {968}},\ \bibinfo {pages}
  {163970} (\bibinfo {year} {2020})},\ \Eprint
  {http://arxiv.org/abs/1912.05243} {arXiv:1912.05243 [physics.ins-det]}
  \BibitemShut {NoStop}%
\bibitem [{\citenamefont {Whitten}(2008)}]{Whitten:2008zz}%
  \BibitemOpen
  \bibfield  {author} {\bibinfo {author} {\bibfnamefont {C.~A.}\ \bibnamefont
  {Whitten}} (\bibinfo {collaboration} {STAR Collaboration}),\ }\bibfield
  {title} {\enquote {\bibinfo {title} {{The beam-beam counter: A local
  polarimeter at STAR}},}\ }\href {\doibase 10.1063/1.2888113} {\bibfield
  {journal} {\bibinfo  {journal} {AIP Conf. Proc.}\ }\textbf {\bibinfo {volume}
  {980}},\ \bibinfo {pages} {390--396} (\bibinfo {year} {2008})}\BibitemShut
  {NoStop}%
\bibitem [{\citenamefont {Adam}\ \emph {et~al.}(2020)\citenamefont {Adam} \emph
  {et~al.}}]{STAR:2019bjj}%
  \BibitemOpen
  \bibfield  {author} {\bibinfo {author} {\bibfnamefont {Jaroslav}\
  \bibnamefont {Adam}} \emph {et~al.} (\bibinfo {collaboration} {STAR
  Collaboration}),\ }\bibfield  {title} {\enquote {\bibinfo {title} {{Strange
  hadron production in Au+Au collisions at $\sqrt{s_{NN}}=$7.7 , 11.5, 19.6,
  27, and 39 GeV}},}\ }\href {\doibase 10.1103/PhysRevC.102.034909} {\bibfield
  {journal} {\bibinfo  {journal} {Phys. Rev. C}\ }\textbf {\bibinfo {volume}
  {102}},\ \bibinfo {pages} {034909} (\bibinfo {year} {2020})},\ \Eprint
  {http://arxiv.org/abs/1906.03732} {arXiv:1906.03732 [nucl-ex]} \BibitemShut
  {NoStop}%
\bibitem [{\citenamefont {Adam}\ \emph {et~al.}(2021)\citenamefont {Adam} \emph
  {et~al.}}]{STAR:2020xbm}%
  \BibitemOpen
  \bibfield  {author} {\bibinfo {author} {\bibfnamefont {Jaroslav}\
  \bibnamefont {Adam}} \emph {et~al.} (\bibinfo {collaboration} {STAR
  Collaboration}),\ }\bibfield  {title} {\enquote {\bibinfo {title} {{Global
  Polarization of $\Xi$ and $\Omega$ Hyperons in Au+Au Collisions at $\sqrt
  {s_{NN}}$ = 200 GeV}},}\ }\href {\doibase 10.1103/PhysRevLett.126.162301}
  {\bibfield  {journal} {\bibinfo  {journal} {Phys. Rev. Lett.}\ }\textbf
  {\bibinfo {volume} {126}},\ \bibinfo {pages} {162301} (\bibinfo {year}
  {2021})},\ \Eprint {http://arxiv.org/abs/2012.13601} {arXiv:2012.13601
  [nucl-ex]} \BibitemShut {NoStop}%
\bibitem [{\citenamefont {Zyzak}()}]{Zyzak:Phd}%
  \BibitemOpen
  \bibfield  {author} {\bibinfo {author} {\bibfnamefont {M.}~\bibnamefont
  {Zyzak}},\ }\href@noop {} {\enquote {\bibinfo {title} {{Online selection of
  short-lived particles on many-core computer architectures in the CBM
  experiment at FAIR}},}\ }\bibinfo {howpublished}
  {\url{https://inis.iaea.org/collection/NCLCollectionStore/_Public/48/031/48031842.pdf}}\BibitemShut
  {NoStop}%
\bibitem [{\citenamefont {Gorbunov}()}]{Gorbunov:Phd}%
  \BibitemOpen
  \bibfield  {author} {\bibinfo {author} {\bibfnamefont {S.}~\bibnamefont
  {Gorbunov}},\ }\href@noop {} {\enquote {\bibinfo {title} {{On-line
  reconstruction algorithms for the CBM and ALICE experiments}},}\ }\bibinfo
  {howpublished} {\url{https://d-nb.info/1043977902/34}}\BibitemShut {NoStop}%
\bibitem [{\citenamefont {Kisel}(2018)}]{Kisel:2018nvd}%
  \BibitemOpen
  \bibfield  {author} {\bibinfo {author} {\bibfnamefont {Ivan}\ \bibnamefont
  {Kisel}} (\bibinfo {collaboration} {CBM}),\ }\bibfield  {title} {\enquote
  {\bibinfo {title} {{Event Topology Reconstruction in the CBM Experiment}},}\
  }\href {\doibase 10.1088/1742-6596/1070/1/012015} {\bibfield  {journal}
  {\bibinfo  {journal} {J. Phys. Conf. Ser.}\ }\textbf {\bibinfo {volume}
  {1070}},\ \bibinfo {pages} {012015} (\bibinfo {year} {2018})}\BibitemShut
  {NoStop}%
\bibitem [{STA()}]{STARnote:644}%
  \BibitemOpen
  \href@noop {} {\enquote {\bibinfo {title} {{STAR Public Note SN0644 -
  Technical Design Report for the iTPC Upgrade}},}\ }\bibinfo {howpublished}
  {\url{https://drupal.star.bnl.gov/STAR/starnotes/public/sn0644}}\BibitemShut
  {NoStop}%
\bibitem [{\citenamefont {Abdallah}\ \emph {et~al.}(2022)\citenamefont
  {Abdallah} \emph {et~al.}}]{STAR:2021mii}%
  \BibitemOpen
  \bibfield  {author} {\bibinfo {author} {\bibfnamefont {Mohamed}\ \bibnamefont
  {Abdallah}} \emph {et~al.} (\bibinfo {collaboration} {STAR}),\ }\bibfield
  {title} {\enquote {\bibinfo {title} {{Search for the chiral magnetic effect
  with isobar collisions at $\sqrt {s_{NN}}$=200 GeV by the STAR Collaboration
  at the BNL Relativistic Heavy Ion Collider}},}\ }\href {\doibase
  10.1103/PhysRevC.105.014901} {\bibfield  {journal} {\bibinfo  {journal}
  {Phys. Rev. C}\ }\textbf {\bibinfo {volume} {105}},\ \bibinfo {pages}
  {014901} (\bibinfo {year} {2022})},\ \Eprint
  {http://arxiv.org/abs/2109.00131} {arXiv:2109.00131 [nucl-ex]} \BibitemShut
  {NoStop}%
\bibitem [{\citenamefont {Goudarzi}\ \emph {et~al.}(2020)\citenamefont
  {Goudarzi}, \citenamefont {Wang},\ and\ \citenamefont
  {Huang}}]{Goudarzi:2020eoh}%
  \BibitemOpen
  \bibfield  {author} {\bibinfo {author} {\bibfnamefont {Amir}\ \bibnamefont
  {Goudarzi}}, \bibinfo {author} {\bibfnamefont {Gang}\ \bibnamefont {Wang}}, \
  and\ \bibinfo {author} {\bibfnamefont {Huan~Zhong}\ \bibnamefont {Huang}},\
  }\bibfield  {title} {\enquote {\bibinfo {title} {{Evidence of coalescence sum
  rule in elliptic flow of identified particles in high-energy heavy-ion
  collisions}},}\ }\href {\doibase 10.1016/j.physletb.2020.135974} {\bibfield
  {journal} {\bibinfo  {journal} {Phys. Lett. B}\ }\textbf {\bibinfo {volume}
  {811}},\ \bibinfo {pages} {135974} (\bibinfo {year} {2020})},\ \Eprint
  {http://arxiv.org/abs/2007.11617} {arXiv:2007.11617 [nucl-ex]} \BibitemShut
  {NoStop}%
\bibitem [{\citenamefont {Riley}\ \emph {et~al.}(2006)\citenamefont {Riley},
  \citenamefont {Hobson},\ and\ \citenamefont {Bence}}]{riley2006mathematical}%
  \BibitemOpen
  \bibfield  {author} {\bibinfo {author} {\bibfnamefont {K.F.}\ \bibnamefont
  {Riley}}, \bibinfo {author} {\bibfnamefont {M.P.}\ \bibnamefont {Hobson}}, \
  and\ \bibinfo {author} {\bibfnamefont {S.J.}\ \bibnamefont {Bence}},\ }\href
  {https://books.google.com/books?id=Mq1nlEKhNcsC} {\emph {\bibinfo {title}
  {Mathematical Methods for Physics and Engineering: A Comprehensive Guide}}}\
  (\bibinfo  {publisher} {Cambridge University Press},\ \bibinfo {year}
  {2006})\BibitemShut {NoStop}%
\end{thebibliography}%

\end{document}